\documentclass{article}
\usepackage{amssymb}
\usepackage{bm}

\usepackage{graphicx}
\usepackage{amsmath}

\def\nsection#1{\section{#1}\setcounter{equation}{0}}

\newcommand{\qq}{\begin{eqnarray}}
\newcommand{\qqq}{\end{eqnarray}}
\newcommand{\ee}{{\rm e}}
\newcommand{\CB}{{\cal B}}
\newcommand{\CC}{{\cal C}}

\newcommand{\CF}{{\cal F}}

\newcommand{\CJ}{{\cal J}}
\newcommand{\CM}{{\cal M}}
\newcommand{\CO}{{\cal O}}
\newcommand{\CR}{{\cal R}}
\newcommand{\CW}{{\cal W}}

\newcommand{\m}{\hspace{0.025cm}}

\begin{document}

\title{\textbf{Fluctuation relations in simple examples}\\
\textbf{of non-equilibrium steady states}}
\author{Rapha$\mathrm{\ddot{e}}$l Chetrite$^{1}$,  
\,Gregory Falkovich$^{2,3}$ \,and \,Krzysztof Gaw\c{e}dzki$^{1}$ \\ 
\\
$^{1}$\small{Laboratoire de Physique, C.N.R.S., ENS-Lyon,
Universit\'e de Lyon, 46 All\'ee d'Italie,}\cr
\hspace*{-8.9cm}\small{69364 Lyon, France}
\hfill\cr
\hspace*{-0.3cm}
$^{2}$\small {Physics of Complex Systems, Weizmann Institute of Science,
Rehovot 76100, Israel}\hfill\cr
\hspace*{-5.49cm}
$^{3}$\small{KITP, UCSB, Santa Barbara, CA
93106, USA}\hfill}
\date{}
\maketitle

\abstract{\noindent We discuss fluctuation relations in simple cases 
of non-equilibrium Langevin dynamics. In particular, we show that close 
to non-equilibrium steady states with non-vanishing probability
currents some of these relations reduce to a modified version of the 
fluctuation-dissipation theorem. The latter may be interpreted as the 
equilibrium-like relation in the reference frame moving with the
mean local velocity determined by the probability current.}
\ 

\vskip 0.5cm
\

\nsection{Introduction}

In statistical mechanics, the fluctuation-dissipation theorem (FDT) provides
a simple relation in an equilibrium state between the response of the 
fixed-time averages to small time-dependent perturbations of the Hamiltonian 
and the dynamical correlations \cite{Kubo0,Kubo}. \,Let $\,O^a(x)\,$
for $\,a=1,\dots,A\,$ be a collection of (classical) observables. With 
the shorthand notation $\,O^a_t\,$ for the single-time functions 
$\,O^a(x_t)\,$ of the dynamical process $\,x_t$, \,the response function 
and the 2-time correlation function in a steady state are, respectively, 
\qq
\CR^{ab}(t-s)\,=\,\frac{\delta}{\delta h_{s}}\Big|_{_{h=0}}\,
\big\langle O^a_t\big\rangle_{\hspace{-0.04cm}_h}\qquad\ {\rm and}\ \ \qquad
\CC^{ab}(t-s)\,=\,\big\langle O^a_t\,O^b_s\big\rangle_{\hspace{-0.03cm}_0}\,,
\label{CRCC}
\qqq
where $\,\langle\,-\,\rangle_{\hspace{-0.03cm}_h}\,$ denotes 
the dynamical expectation obtained from the steady state by
replacing the time-independent Hamiltonian $\,H(x)\,$ by a slightly
perturbed time-dependent one $\,H(x)-h_{t}\m O^{b}(x)$.
\,The FDT asserts that when the unperturbed state is the equilibrium 
at inverse temperature $\,\beta\,$ then
\qq
\beta ^{-1}\,\CR^{ab}(t-s)
\ =\ \partial_s\,\CC^{ab}(t-s)\,.
\label{EFDT}
\qqq 
Such a direct relation between the response and correlation functions is 
violated in systems out of equilibrium and a lot of interest in the 
research on non-equilibrium statistical mechanics was devoted to such 
violations. In particular, they were studied intensively for glassy systems 
\cite{CK,Cri,Cal}, for colloidal suspensions \cite{Bar,Fuc}, for granular 
matter \cite{BCL,BuShLe}, and for biophysical systems \cite{Mar,HayTak}. 
In recent years, it has been realized that the FDT, as well as the 
Green-Kubo relation, another linear response law of the equilibrium regime, 
are special cases of more general fluctuation relations that hold also 
far from equilibrium. Such relations pertain either to non-stationary 
transient situations \cite{Eva,Jarz} or to stationary regimes \cite{Gal}. 
In particular, the so called Jarzynski equality\cite{Jarz} for the dynamics 
with a time dependent Hamiltonian reduces to the FDT for tiny time 
variations \cite{Che}.
\vskip 0.1cm

In the present paper, we revisit the violations of the FDT in simple 
examples of non-equilibrium steady states (NESS) for systems 
with few degrees of freedom evolving according to the Langevin equation 
possibly including non-conservative forces, see e.g. 
\cite{Risk,CKP,MPRV,HaySas,HarSas,SpS0}. For such systems, we show 
a modified fluctuation-dissipation theorem (MFDT) that may be written 
in the form
\qq 
\beta ^{-1}\CR^{ab}_{\hspace{-0.02cm}_L}(t,s)\ =\ \partial_s\,
\CC^{ab}_{\hspace{-0.02cm}_L}(t,s)
\label{extend}
\qqq 
similar to the equilibrium relation, somewhat 
in the spirit of ref.\,\cite{SpS0}. Above, $\,\CR^{ab}_{_L}(t,s)\,$ 
and $\,\CC^{ab}_{_L}(t,s)\,$ denote the response function and the dynamical 
correlation function in the 
Lagrangian frame moving with the mean local 
velocity $\,\nu_{\hspace{-0.01cm}_0}(x)\,$ of the NESS. \,The Lagrangian-frame 
functions are obtained by replacing in the definitions (\ref{CRCC}) the 
time-independent observables $\,O^a(x)\,$ by the time-dependent ones 
$\,O^a(t,x)\,$ that evolve according to the advection equation
\qq
\partial_t\,O^a(t,x)\,+\,\nu_{\hspace{-0.01cm}_0}(x)\cdot\nabla\,
O^a(t,x)\,=\,0\,,
\label{adv}
\qqq
i.e.\ are frozen in the Lagrangian frame. In the equilibrium, the mean 
local velocity $\,\nu_{\hspace{-0.01cm}_0}(x)\,$ vanishes and the MFDT 
(\ref{extend}) becomes the FDT (\ref{EFDT}). 
\vskip 0.1cm
 
The other goal of the present work is to explain how the MFDT 
(\ref{extend}) may be obtained from more general fluctuation relations 
by restricting them to the regime close to NESS, \,similarly as 
for the case of the equilibrium FDT. \,Before doing that, we recall
different fluctuation relations holding arbitrarily far from stationarity
and equilibrium in Langevin systems and their perturbations.
Our discussion follows with minor modifications the recent 
exposition \cite{Che}.   
\vskip 0.1cm

The general results presented in the paper apply, in particular, 
to two types of one-dimensional systems with NESS. The first type 
describes an exploding Langevin dynamics on the line with a non-Gibbsian 
invariant measure. This case arises, for example, when 
one studies the tangent process for particles with inertia moving in 
one-dimensional Kraichnan's random velocities \cite{Kr68,FGV}. Such 
velocities $\,v_t(x)\,$ form a Gaussian ensemble with mean zero and 
covariance
\begin{equation}
\big\langle v_t(x)\,v_s(y)\big\rangle =\ \delta (t-s)\,D(x-y)\,.
\end{equation}
The evolution of the inertial particles is described by the stochastic
differential equation (SDE) \cite{Bec}
\begin{equation}
\dot{x}\ =\ u\,,\qquad\dot{u}\ =\ \frac{_{1}}{^{\tau }}(-u+v_t(x))\,,
\label{inert}
\end{equation}
where $\,\tau \,$ is the Stokes time measuring the time-delay of
the particles relative to the flow motion. The separation between two
infinitesimally close trajectories satisfies then the equations
\begin{equation}
\frac{_{d}}{^{dt}}\delta x\ =\ \delta u\,,\qquad\frac{_{d}}{^{dt}}\delta u\
=\ \frac{_{1}}{^{\tau }}\big(-\delta u+\delta x\m\,\partial_xv_t(x)\big)\,,
\end{equation}
where one may replace $\,\frac{_{1}}{^{\tau }}\partial_{x}v_t(x)\,$ on the
right hand side by a white noise $\,\zeta_t\,$ with the covariance
\begin{equation}
\big\langle \zeta_t\,\zeta_s\big\rangle \,=\ 2\beta^{-1}\delta(t-s)
\label{cova}
\end{equation}
for $\,\beta^{-1}=-\frac{1}{2\tau^2}\partial_x^2D(0)$.
\,For the ratio $\,X=\frac{\delta u}{\delta x}$, \,one obtains the SDE
\begin{equation}
\dot{X}\ =\ -X^{2}\,-\,\frac{_{1}}{^{\tau }}X\,+\,\zeta\ =\
-\partial_{\hspace{-0.01cm}_X}H(X)\,+\,\zeta  \label{SDEX}
\end{equation}
which has the form of a one-dimensional overdamped Langevin equation for
the Hamiltonian $\,H(X)=\frac{1}{3}X^{3}+\frac{1}{2\tau }X^{2}$. \,The process
$\,X_t\,$ solving Eq.\,(\ref{SDEX}) escapes in finite time to $\,-\infty\,$ 
but has a realization with trajectories that reappear immediately from 
$\,+\infty$, \,see Appendix A. \,This corresponds to the solutions for
$\,(\delta x, \delta u)\,$ where $\,\delta
x\, $ passes through zero with a non-vanishing speed, i.e.\,\,to the crossing of
close particle trajectories with faster particles overcoming slower ones
(allowed in this model of a dilute particle suspension with no pressure 
and no back-reaction on the flow \cite{DFTT}). 
The Gibbs density $\,\ee^{-\beta H}\,$ is not normalizable here. 
The resurrecting process has, however, a non-Gibbsian
invariant probability measure with constant probability flux.
The top Lyapunov exponent for the random dynamical system (\ref{inert})
is obtained as the mean value in this measure of $\,X\,$ (which is the
temporal logarithmic derivative of $\,|\delta x|$)  \cite{Wilk}.
The above is a variation of a much older story \cite{Halp,LGP} of the 
one-dimensional Anderson localization in the stationary Schr\"odinger 
equation
\begin{equation}
-\frac{_{d^2}}{^{dx^2}}\psi\,+\,V\,\psi\ =\ E\,\psi
\end{equation}
with a $\,\delta$-correlated potential $\,V(x)$. \,For $\,Y=(\frac{d}{dx}\psi)
/\psi$, \,one obtains the equation
\begin{equation}
\frac{_{d}}{^{dx}}Y\ =\ -Y^{2}\,-\,E\,+\,V  \label{SDEY}
\end{equation}
that may be viewed as a stochastic evolution equation
if $\,x\,$ is interpreted as time. The invariant measure with constant
flux for such an SDE was already used in \cite{Halp}. The substitution
$\,Y=X+\frac{1}{2\tau}$, $\,E=-\frac{1}{4\tau^2}$, $\,V=\zeta\,$ turns
Eq.\,(\ref{SDEY}) to Eq.\,(\ref{SDEX}) (provided that one
replaces $\,x\,$ by $\,t$). 
\vskip 0.1cm

The second particular type of systems with NESS covered by our discussion
is obtained by adding a non-conservative force to the Langevin dynamics. 
More specifically, we shall consider a particle which moves on a  
circle according to the SDE
\begin{equation}
\dot{x}\ =\ -\partial _{x}H(x)+G(x)\,+\,\zeta\,,  \label{1dLan}
\end{equation}
where, as before, $\,\zeta_t\,$ is the white noise with covariance (\ref
{cova}). The above dynamics pertains again to the overdamped regime where it
is the particle velocity rather than the particle acceleration that is
proportional to the force. The angular coordinate $\,x\,$ will be taken
modulo $\,2\pi$. \,We
shall assume that $\,H(x+2\pi)=H(x)\,$ and $\,G(x+2\pi)=G(x)\,$ but
$\,\int_0^{2\pi}G(x)dx\not=0\,$ so that the force $\,G\,$ is not a gradient
and it drives the system out of equilibrium. \,Eq.\,(\ref{1dLan}) was used,
for example, to describe the motion of a colloidal particle in an optical 
trap \cite{SBBS}. It was discussed recently in \cite{Maes} in a
context similar to the one of this work.
\vskip 0.1cm

The present paper is organized as follows. Sec.\,2, returns to the
discussion of stationary Langevin diffusion processes, presenting more
details on the one-dimensional systems with explicit non-Gibbsian
invariant measures \cite{Che,Maes}. For such systems, we examine 
in Sec.\,3 the simplest fluctuation-response relation that describes 
the change of the invariant measure under a small time-independent 
variation of the Hamiltonian. In Sec.\,4, we prove the MFDT (\ref{extend}) 
that holds around NESS of the Langevin-type dynamics, in particular, 
in the one-dimensional cases with explicit invariant measures.
Sec.\,5 is devoted to a brief presentation of general fluctuation 
relations for SDE's \cite{Kurchan,LebowSp,HatSas,Kurchan1,Che}. These 
are specified for the Langevin systems under consideration in Sec.\,6.
In particular, we describe the Crooks detailed fluctuation relation
\cite{Crooks2} and the Hatano-Sasa \cite{HatSas} version
of the Jarzynski equality \cite{Jarz1,Jarz2}, both holding arbitrarily 
far from stationarity. In Sec.\,7, we return to the MFDT, showing that 
it may be viewed as a limiting case around the stationary situation 
of the Crooks transient fluctuation relation or, in a special case, 
of the Jarzynski-Hatano-Sasa equality. Finally, after brief Conclusions, 
we collect in Appendix A some facts about the one-dimensional processes 
with explosion and illustrate in Appendix B the MFDT by an explicit
calculation for the Langevin particle driven by a constant 
force along a circle.
\vskip 0.2cm

\noindent{\bf Acknowledgements}. \ The work of G.F. was supported 
in part by the US National Science Foundation under Grant No. PHY05-51164 
and by the Minerva Foundation.

\nsection{NESS in Langevin processes}

The general stationary dynamics that we consider is described by the
Langevin equation in $d$-dimensions with an external force:
\begin{equation}
\dot{x}^i\ =\
-\Gamma^{ij}\partial_{j}H(x)+\Pi^{ij}\partial_{j}H(x)\,+\,G^i(x) +\,\zeta^i\,,
\label{sys}
\end{equation}
where $\,\Gamma\,$ is a constant non-negative matrix and $\,\Pi\,$ an
antisymmetric one, $\,H\,$ is the Hamiltonian, $\,G\,$ the external
force and the white noise $\,\zeta_t\,$ has the covariance
\begin{equation}
\big\langle \zeta^i_t\,\zeta^j_s\big\rangle =2\beta
^{-1}\Gamma^{ij}\delta (t-s)\,.  \label{LEq}
\end{equation}
The deterministic force $-\Gamma \nabla H$ decreases the energy,
driving the solution towards the minimum of $\,H$, \,if it exists,
whereas the noise mimics the effect of a thermal bath.
\,We have added the Hamiltonian force $\Pi \nabla H\,$
that preserves the energy in order to cover systems governed 
by Langevin-Kramers equations \cite{Kurchan} or Fermi-Pasta-Ulam 
chains \cite{Eckm}. \,The generator $\,L\,$ of the process 
$\,x_t\,$ satisfying the SDE (\ref{sys}) is defined by the relation
\qq
\partial_t\big\langle f(x_t)\big\rangle\,=\,\big\langle(Lf)(x_t)\big\rangle\m.
\qqq
It is a second order differential operator:
\qq
L\,=\,\big((-\Gamma+\Pi)(\nabla H)+G\big)\cdot\nabla\,
+\,\beta^{-1}\nabla\cdot\Gamma\nabla
\label{L}
\qqq
in the vector notation, \,with the formal adjoint
\qq
L^\dagger\,=\,-\,\nabla\cdot\big((-\Gamma+\Pi)(\nabla H)+G\big)\,+\,
\beta^{-1}\nabla\cdot\Gamma\nabla\,.
\label{Ldag}
\qqq
The transition probabilities
$\,P_t(x,dy)\,$ of the process satisfy the evolution equation
\qq
\partial_t\,P_t(x,dy)\,=\,L_x\,P_t(x,dy)\,,
\label{prev}
\qqq
with the subscript in $\,L_x\,$ indicating that $\,L\,$ acts on the
variable $\,x$. \,The dynamics of the mean instantaneous 
density $\,\varrho_t\,$ of the process $\,x_t\,$
is generated by the adjoint operator $\,L^\dagger\,$ and takes the form
of the continuity equation
\begin{eqnarray}
\partial_t\varrho_t\,=\,L^\dagger\m\varrho_t\,=\,-\,\nabla\cdot j_t
\label{coneq}
\end{eqnarray}
with the current\footnote{For convenience, we have included
into the current the term $\,\beta^{-1}\Pi\nabla\varrho$.} 
\qq
j_t\,=\,\big(-\Gamma(\nabla H)+\Pi(\nabla H)\m+G-
\beta^{-1}(\Gamma-\Pi)\nabla\big)\,\varrho_t\,.
\label{curr0}
\qqq
Following \cite{HatSas}, let us introduce the mean local velocity 
$\,\nu_t=\varrho_t^{-1}j_t$. \,With the use of the velocity field, 
the above continuity equation may be rewritten in the hydrodynamical 
form as the advection equation for the density $\,\varrho_t\m$,
\qq
(\partial_t\,+\,\nabla\cdot\nu_t)\varrho_t\,=\,0\,,
\label{conder}
\qqq
stating that $\,\rho_t\,$ is annihilated by the convective derivative
or, in other words, that it evolves as the density of Lagrangian
particles whose trajectories obey the ordinary differential equation 
\qq
\dot x\,=\,\nu_t(x)\,.
\label{LTr}
\qqq 
\vskip 0.1cm

For an invariant density, \,the corresponding current is conserved: 
$\,\nabla\cdot j=0$. \,If for the density $\,\varrho\,$  
the current $\,j\,$ itself vanishes then one says that the dynamics 
(\ref{LEq}) satisfies the detailed balance relative to $\,\varrho$. 
\,The detailed balance holds relative to the 
Gibbs density $\,\ee^{-\beta H}\,$ if $\,G=0\,$
(this was assured by the addition of the term 
$\,\beta^{-1}\Pi\nabla\varrho\,$ to the current). 
\,When $\,G\not=0$, \,the invariant density is not known 
explicitly, in general, even if it exists. There are, however, special 
cases of processes satisfying the SDE (\ref{LEq}) where one may obtain an
analytic formula for a non-Gibbsian invariant density. 

\subsection{NESS for resurrecting processes}

The first example,
that we shall call type 1 below, is obtained for the Langevin equation
on the line. In this case, any force is a gradient so that, upon setting
for simplicity $\,\Gamma=1$, the dynamical equation becomes
\begin{equation}
\dot{x}\ =\ -(\partial _{x}H)(x)\,+\,\zeta  \label{SDEx}
\end{equation}
with the covariance of the white noise given by Eq.\,(\ref{cova}). The detailed
balance holds here relative to the Gibbs density
$\,\mathrm{e}^{-\beta H}\,$ since the corresponding current vanishes.
\,Such a density may, however, be not normalizable, hence not
leading to an invariant probability measure. Let us look closer at various
possibilities by considering the case of a polynomial Hamiltonian with
the highest degree term equal to $\,ax^k\,$ \cite{Che}.
\,For $\,k=0\,$ or $\,k=1$, \,the solution $\,x_t\,$ of Eq.\,(\ref{SDEx})
is, up to a linear change of variables, a Brownian motion, which does
not have an invariant probability measure. For
even $\,k\geq 2\,$ and $\,a>0$, \,the Gibbs measure
$\,\mu(dx)
=Z^{-1}\mathrm{e}^{-\beta H(x)}dx\,$ provides the unique invariant probability
measure of the process $x_t$. \,If $\,a<0$, however, then the Gibbs density
is not normalizable. In this case the process escapes with probability one
to $\,\pm \infty \,$ in a finite time and it has no invariant probability
measure. \,For odd $\,k\geq 3$, \,the Gibbs density $\mathrm{e}^{-\beta
H(x)}\,$ is also not normalizable. Changing eventually $\,x\,$ to $\,-x$,
we may assume that $\,a>0$. \,In this case, the process $\,x_t\,$ escapes
in finite time to $\,-\infty\,$ but it has a realization with the
trajectories that reappear immediately from $\,+\infty$, \,as discussed
in Appendix A. \,Such a resurrecting process has a unique invariant
probability measure
\begin{equation}
\mu(dx)\,=\,\frac{1}{Z}\,\Big(\int\limits_{-\infty
}^{x} \mathrm{e}^{\hspace{0.025cm}\beta H(y)}dy\Big)\,
\mathrm{e}^{-\beta H(x)}\,dx\ \equiv\
\varrho_{\hspace{-0.01cm}_H}\hspace{-0.01cm}(x)\,dx\,,  \label{C1}
\end{equation}
where $\,Z\,$ is the (positive) normalization constant. The density
$\,\varrho_{\hspace{-0.01cm}_H}\,$ of $\,\mu\,$ behaves as
$\,(Za\beta k x^{k-1})^{-1}\,$ when
$\,x\to\pm\infty$, see the estimate (\ref{estim}) in Appendix A.
\m It corresponds to the constant current $\,j=-(\beta
Z)^{-1}\,$ with the flux towards negative $\,x$. \,The situation provides
one of the simplest examples of
NESS. \,In particular, the inertial particle in the one-dimensional
Kraichnan flow and the Anderson localization in the one-dimensional
$\,\delta$-correlated potential lead naturally to the resurrecting
processes corresponding to $\,k=3$, as discussed in the introduction.

\subsection{NESS for forced diffusions on circle}

The second model with an explicit analytic expression for the invariant
non-Gibbsian measure, that we shall call type 2 below, is the perturbed
one-dimensional Langevin equation (\ref{1dLan}) on the unit circle. Now,
the unique invariant probability measure is given by the formula
\cite{Risk,HaySas,Maes}
\begin{equation}
\mu(dx)\,=\,\frac{1}{Z}\Big(\int\limits_{0}^{2\pi}
\mathrm{e}^{\m\beta
U(x,y)}dy\Big)\,\ee^{-\beta H(x)}\,dx \ \equiv\ \varrho_{\hspace{-0.01cm}_H}
\hspace{-0.01cm}(x)\,dx  \label{C2}
\end{equation}
for $\,0\leq x\leq 2\pi$, \,\,with
\begin{equation}
U(x,y)\,=\,H(y)\,+\,\theta(x-y)\int\limits_{y}^{x}G(z)\,dz\,
+\,\theta (y-x)\Big(\int\limits_{0}^{x}G(z)\,dz
\m+\int\limits_{y}^{2\pi}G(z)\,dz\Big).
\end{equation}
Also here the density $\,\varrho_{\hspace{-0.01cm}_H}\,$ of
the measure $\mu\,$ corresponds to a constant probability current
\begin{equation}
j\,=\,\frac{1}{\beta Z}\Big(\ee^{\,\beta \int\limits_{0}^{2\pi}G(z)\,dz}-\,1
\Big).
\end{equation}
In the following, we shall see how the presence of the probability 
flux in NESS deforms the usual fluctuation relations.

\nsection{Modified fluctuation-response relation}

As a warmup, let us see what is the form taken by the most elementary
fluctuation-response relation \cite{LeBellac} in the one-dimensional
systems with NESS that we discussed above. The setup of the   
fluctuation-response relation is as follows. One prepares the system in 
the far past in the invariant state with probability density
$\,\varrho_{\hspace{-0.01cm}_H}\,$ that is given by
Eqs.\,(\ref{C1}) and (\ref{C2}) for the type 1 and type 2 systems,
respectively. \,At $\,t=0$, \,the Hamiltonian $\,H\,$ is perturbed by a
small time-independent potential $\,V\,$ (vanishing sufficiently
fast when $\,x\to\pm\infty\,$ for type 1), leading to the change
$\,H\mapsto H^{\prime}=H+V$.
\,The systems\ evolves then and converges toward the new steady state
with the probability density $\,\varrho_{\hspace{-0.01cm}_{H^{\prime}}}$.
\,The fluctuation-response relation compares the initial and the final
averages of an observable $\,O_t\equiv O(x_t)\m$:
\begin{eqnarray}
\left\langle O_{\hspace{-0.02cm}_0}\right\rangle\,
=\,\int O(x)\,\varrho_{\hspace{-0.01cm}_H}(x)\,dx\qquad
{\rm and}\qquad
\left\langle O_{\hspace{-0.03cm}_\infty}\right\rangle\,
=\,\int O(x)\,\varrho_{\hspace{-0.01cm}_{H^{\prime}}}(x)
\,dx\,.
\end{eqnarray}
By a straightforward differentiation of the explicit formulae for the
invariant densities, one obtains the identity
\begin{equation}
\big\langle O_{\hspace{-0.03cm}_\infty}\big\rangle\,
=\,\big\langle O_{\hspace{-0.02cm}_0}
\big\rangle-\beta\big[\big\langle O_{\hspace{-0.02cm}_0}
(V_{\hspace{-0.02cm}_0}-\widehat{V}_{\hspace{-0.02cm}_0})
\big\rangle-\big\langle O_{\hspace{-0.02cm}_0}
\big\rangle\big\langle V_{\hspace{-0.02cm}_0}
-\widehat{V}_{\hspace{-0.02cm}_0}\big\rangle\big]\,,
\label{FRR}
\end{equation}
up to terms of the second order in $\,V$, \,where
\begin{equation}
\widehat{V}(x)\,=\,\frac{\int\limits_{-\infty }^{x}V(y)\, \mathrm{e}^{
\hspace{0.025cm}\beta H(y)}dy}{\int\limits_{-\infty }^{x}\,\mathrm{e}^{
\hspace{0.025cm}\beta H(y)}dy}\ \qquad\mathrm{and}\ \qquad \widehat{V}(x)\,=\,
\frac{\int\limits_{0}^{2\pi}V(y)\,\mathrm{e}^{\m\beta U(x,y)}dy} {%
\int\limits_{0}^{2\pi}\mathrm{e}^{\m\beta U(x,y)}dy}
\label{hat12}
\end{equation}
for the systems of type 1 and type 2, respectively. \,Eq.\,(\ref{FRR})
deforms the usual fluctuation-response relation around the Gibbs state
\cite{LeBellac} by replacing $\,V\,$
by $\,(V-\widehat{V})\,$ with the averaged potential $\,\widehat{V}\,$
dependent on the initial Hamiltonian $\,H$.

\nsection{Modified fluctuation-dissipation theorem}

Coming back to the general case, let us consider a system prepared at negative 
times in the steady state of the stationary Langevin dynamics 
(\ref{sys}). This forces the time zero value $\,x_{\hspace{-0.01cm}_0}\,$ 
of the corresponding process to be distributed according to the invariant 
probability measure $\,\mu_{\hspace{-0.01cm}_0}(dx)
=\varrho_{\hspace{-0.01cm}_0}(x)\,dx$.
\,At \,$t=0$, \,one switch on a non-stationary perturbation taking 
the Hamiltonian for the positive times to be equal to
\begin{equation}
H_{t}(x)\,=\,H(x)-\sum\limits_{a}h_{a,t}\m O^{a}(x)\,,
\label{hpert}
\end{equation}
where $\,h_{a,t}\,$ carry the time dependence
and functions $\,O^a(x)\,$ (the ``observables'') are supposed,
for simplicity, to vanish sufficiently fast when $\,|x|\to\infty$.
\,We denote by $\,\big\langle\CF\big\rangle_{\hspace{-0.03cm}_h}\,$
the corresponding expectation, with $\,\big\langle\CF
\big\rangle_{\hspace{-0.03cm}_0}\,$
referring to the non-perturbed case. The expression
\qq
\big\langle\CF\,R^a_t\big\rangle_{\hspace{-0.03cm}_0}\,=\,
\frac{\delta}{\delta h_{a,t}}\Big|_{h=0}\,\big\langle\CF\big\rangle\,.
\label{resp}
\qqq
defines the response correlations. \,To shorten further the notations,
let us set
\qq
\big\langle O^a_t\,R^b_s\big\rangle_{\hspace{-0.03cm}_0}\ \equiv\ \CR^{ab}(t-s)
\,,\qquad\big\langle O^a_t\,O^b_s\big\rangle_{\hspace{-0.03cm}_0}\ \equiv\
\CC^{ab}(t-s)\,,\qquad
\theta(t-s)\,\big\langle O^a_t\,B^b_s\big\rangle_{\hspace{-0.03cm}_0}\ \equiv\
\CB^{ab}(t-s)
\qqq
for $\,O_t\equiv O(x_t)\,$ and the induced observable 
\qq
B^b\,=\,\varrho_{\hspace{-0.01cm}_0}^{-1}j_{\hspace{-0.01cm}_0}\cdot\nabla O^b
\label{Bb}
\qqq
with $\,j_{\hspace{-0.01cm}_0}\,$ standing for the current
(\ref{curr0}) corresponding to the invariant density
$\,\varrho_{\hspace{-0.02cm}_0}$. \,Note that for the one-dimensional NESS 
with constant probability current $\,j_{\hspace{-0.01cm}_0}$, \,the observable
$\,B^b=j_{\hspace{-0.01cm}_0}\m\varrho_{\hspace{-0.01cm}_0}^{-1}
\partial_x O^b\,$ has the probability flux as an explicit factor.
Remark that, by causality, the response function
$\,\CR^{ab}(t-s)\,$ vanishes for $\,s\geq t$.
\,We shall show the following modified version of the
fluctuation-dissipation theorem (MFDT) holding for $\,t>s\m$:
\qq
\beta^{-1}\,\CR^{ab}(t-s)\ =\ \partial _{s}\,\CC^{ab}(t-s)\,
-\,\CB^{ab}(t-s)\,.
\label{mfd}
\qqq
The second term on the right hand side of Eq.\,(\ref{mfd})
is new as compared to the FDT around the equilibrium steady state.
Indeed, in the equilibrium case, the external force
$\,G=0\,$ and $\,\varrho_{\hspace{-0.01cm}_0}
=Z^{-1}\ee^{-\beta H}\,$ is the  normalized
Gibbs factor with $\,j_{\hspace{-0.01cm}_0}=0\,$ so that 
$\,B^b=0\,$ and the MFDT (\ref{mfd})
reduces to the standard equilibrium form (\ref{EFDT}).

\subsection{Lagrangian-frame interpretation}

What the formula (\ref{mfd}) for the response function means 
is better understood by rewriting it with the explicit form
of the right hand side as
\qq
\beta^{-1}\,\CR^{ab}(t-s)\hspace{-0.2cm}&=&\hspace{-0.3cm}
\int\hspace{-0.05cm}dx\,\,O^b(x)\,\varrho_{\hspace{-0.01cm}_0}
(x)\,\,\partial_s\m P_{t-s}(x,dy)
\,\,O^a(y)\,-\int\hspace{-0.1cm}dx\,\,(j_{\hspace{-0.01cm}_0}
\cdot\nabla O^b)(x)\,\,P_{t-s}(x,dy)\,\,O^a(y)\qquad\label{rmfd}\\
&=&\hspace{-0.3cm}\int\hspace{-0.05cm}dx\,\,O^b(x)
\,\,\big(\partial_s+\nabla\cdot\nu_{\hspace{-0.01cm}_0}(x)\big)
\hspace{-0.05cm}\int\hspace{-0.05cm}
\varrho_{\hspace{-0.01cm}_0}(x)\,P_{t-s}(x,dy)\,\,O^a(y)\,,\quad
\nonumber
\qqq
where $\,P_t(x,dy)\,$ denotes the stationary transition probabilities
of the unperturbed process and we have integrated by parts to obtain
the second equality, setting $\,\nu_{\hspace{-0.01cm}_0}
=\varrho_{\hspace{-0.01cm}_0}^{-1}j_{\hspace{-0.01cm}_0}$.
\,Note that the time derivative $\,\partial_s\,$ of the equilibrium 
relation has been replaced by the convective derivative 
$\,\partial_s+\nabla\cdot\nu_{\hspace{-0.01cm}_0}(x)\,$  
which acts on the first component of the joint probability 
density function of the time $\,s\,$ and time $\,t\,$ values of 
the stationary process $\,x_t$. \,This suggests that the MFDT 
(\ref{mfd}) should take the equilibrium form in the Lagrangian 
frame moving with the stationary mean local velocity 
$\,\nu_{\hspace{-0.01cm}_0}(x)$. 
\vskip 0.1cm

To render this interpretation more transparent, let us replace 
the time-independent observables $\,O^a(x)\,$ by the time dependent 
ones $\,O^a(t,x)\,$ evolving according to the advection equation (\ref{adv}).
We shall define the Lagrangian-frame response function and correlations
function by
\qq
\CR_{_L}^{ab}(t,s)\,=\,\frac{\delta}{\delta h_{b,s}}
\Big|_{_{h=0}}\,\big\langle\,O^a_t(t)
\big\rangle_{\hspace{-0.03cm}_h}\,,
\qquad\ \CC_{_L}^{ab}(t,s)\,=\,\big\langle O^a_t(t)\,
O^b_s(s)\big\rangle_{\hspace{-0.03cm}_0}\,,
\qqq
where $\,\big\langle\,-\,\big\rangle_{\hspace{-0.03cm}_h}\,$
denotes now the expectation referring to the Hamiltonian
$\,H_t(x)=H(x)-\sum\limits_ah_{t,a}\,O^a(t,x)$. 
$\,O_t(t)\equiv O(t,x_t)\,$ with the double time-dependence. Writing 
the MFDT (\ref{mfd}) with the explicit right hand side given 
by Eq.\,(\ref{rmfd}) for the observables $\,O^a(t,x)$, \,one casts 
this relation into the form
\qq
\beta^{-1}\,\CR^{ab}_{_L}(t,s)&=&
\int dx\,\,O^b(s,x)\,\varrho_{\hspace{-0.01cm}_0}
(x)\,\,\partial_s\m P_{t-s}(x,dy)
\,\,O^a(t,y)\cr
&&-\int dx\,\,\varrho_{\hspace{-0.01cm}_0}(x)\,
\big(\nu_{\hspace{-0.01cm}_0}(x)\cdot\nabla O^b(s,x)\big)\,\,
P_{t-s}(x,dy)\,\,O^a(t,y)\cr
&=&\partial_s\int dx\,\,O^b(s,x)\,\varrho_{\hspace{-0.01cm}_0}
(x)\,\,P_{t-s}(x,dy)
\,\,O^a(t,y)\ =\ 
\partial_s\,\CC^{ab}_{_L}(t,s)\,,
\qqq 
where the last equality follows from the advection equation 
(\ref{adv}). This proves the identity (\ref{extend}) 
announced in the introduction. We should stress that, in spite of
the similarity between that relation and the equilibrium FDT
(\ref{EFDT}), in general it is not true that the dynamical process
$\,x_t\,$ viewed in the Lagrangian frame of the velocity field $\,\nu_0\,$
is governed by an equilibrium Langevin equation, although this is 
what happens in the simple example considered in Appendix B. 
\vskip 0.1cm

For the Langevin process on the circle, 
a fluctuation-dissipation relation for velocities similar to (\ref{mfd}) 
was discussed in \cite{SpS0}, see Eq.\,(11) there, with the interpretation 
similar in spirit, but not in form, to the above one, see the 
subsequent discussion there. One of the consequences of the 
fluctuation-dissipation relation of \cite{SpS0} linking the
effective diffusivity and mobility was checked 
experimentally in \cite{BSLSB}, see also \cite{HayTak0}.
\vskip 0.1cm

It is sometimes more interesting, especially for applications, to 
re-express the fluctuation-dissipation relations in terms of the 
frequency space quantities. Let
\qq
&&\hspace*{-0.9cm}\hat\CC^{ab}(\omega)\,=\,\int\limits_{-\infty}^{\infty}
\ee^{i\m\omega(t-s)}\,\CC^{ab}(t-s)\,dt\,=\,\int\limits_{s}^{\infty}
\ee^{i\m\omega(t-s)}\,\CC^{ab}(t-s)\,dt\,+\,\int\limits_{s}^{\infty}
\ee^{-i\m\omega(t-s)}\,\CC^{ba}(t-s)\,dt\,,\quad\label{C}\\
&&\hspace*{-0.9cm}\hat\CR^{ab}(\omega)\,=\,\int\limits_{s}^{\infty}
\ee^{i\m\omega(t-s)}\,\CR^{ab}(t-s)\,dt\,,\qquad\qquad\qquad\qquad\m\,
\hat\CB^{ab}(\omega)\,=\,
\int\limits_{s}^{\infty}
\ee^{i\m\omega(t-s)}\,\CB^{ab}(t-s)\,dt\,.
\label{RB}
\qqq
Note that $\,\hat\CR^{ab}(\omega)\,$ measures the response to the
time-dependent potential of frequency $\,\omega$.
\,The MFDT (\ref{mfd}) is equivalent to the relation
\qq
i\m\omega\m\,\hat\CC^{ab}(\omega)\ =\ \beta^{-1}\big(\hat\CR^{ab}(\omega)-
\hat\CR^{ba}(-\omega)\big)\,+\,\hat\CB^{ab}(\omega)\,-\,\hat\CB^{ba}(-\omega)
\label{Fmfd}
\qqq
in the frequency space.
\vskip 0.1cm

In general, assuming that the transition probabilities $\,P_{t}(x,dy)\,$ 
converge at long times to the invariant measure, all three terms of 
Eq.\,(\ref{mfd}) tend to zero when $\,(t-s)\to\infty$. \,Mimicking the 
idea employed with success for disordered systems \cite{CK,CKP,Cal}, 
their relative proportions, or the relative proportions of the corresponding 
terms in Eq.\,(\ref{Fmfd}), could be used to define dynamical temperatures 
that would, in general, depend also on the observables involved. We discuss 
those proportions in a simple case of the Langevin dynamics on a circle 
with a constant force in Appendix B. 
\vskip 0.1cm

We propose three derivations of the result (\ref{mfd}): the first
one direct, that we shall present now, and the next two ones
from the general fluctuation relations that will be discussed
in the subsequent sections.

\subsection{Direct derivation}

The beginning of the argument is quite standard, see e.g.\ \cite{Agarw}
or Sec. 2.3.2 of \cite{MPRV}. \,By the definition of the response 
correlations,
\begin{equation}
\CR^{ab}(t-s)\
=\ \frac{\delta }{\delta h_{b,s}}\Big|_{h=0}\,\int O^{a}(y)\,\varrho
_{t}(y)\,dy\,,
\end{equation}
where $\,\varrho _{t}\,$ is the density obtained by the perturbed dynamical
evolution (\ref{coneq}) from $\,\varrho_{\hspace{-0.01cm}_0}$.
\,Using the explicit
form (\ref{curr0}) of the current, one obtains by the first order
perturbation the relation
\begin{equation}
\varrho _{t}(y)\,dy\,=\,\varrho_{\hspace{-0.01cm}_0}(y)\,dy\,
-\,\sum\limits_b\int\limits_0^t h_{b,s}\,ds
\int dx\,\big[\nabla\cdot\big((\Gamma-\Pi)(\nabla O^b)\,
\varrho_{\hspace{-0.01cm}_0}\big)\big](x)
\,\,P_{t-s}(x,dy)\ +\ \CO(h^2)\,.\ 
\end{equation}
Consequently, for $\,t>s$,
\qq
\CR^{ab}(t-s)\ =\ -
\int dx\,\big[\nabla\cdot\big((\Gamma-\Pi)(\nabla O^b)\,
\varrho_{\hspace{-0.01cm}_0}\big)\big](x)
\,\,P_{t-s}(x,dy)\,\,O^a(y)\,.
\label{resp1}
\qqq
Now, a straightforward although somewhat tedious algebra shows that
\qq
\beta^{-1}\nabla\cdot\big((\Gamma-\Pi)(\nabla O^b)\,\varrho_{\hspace{-0.01cm}_0}\big)
\ =\ L^\dagger(O^b\varrho_{\hspace{-0.01cm}_0})\,
+\,\nabla\cdot(O^b j_{\hspace{-0.01cm}_0})
\ =\ L^\dagger(O^b\varrho_{\hspace{-0.01cm}_0})\,
+\,j_{\hspace{-0.01cm}_0}\cdot\nabla O^b\,,
\label{fstal}
\qqq
where the adjoint generator $\,L^\dagger\,$ is given by Eq.\,(\ref{Ldag}) and
in the last equality we have used the conservation of the current
$\,j_{\hspace{-0.01cm}_0}$. \,Substituting this identity to Eq.\,(\ref{resp1})
and integrating the term with $\,L^\dagger\,$ by parts, we obtain
the relation
\qq
\beta^{-1}\,\CR^{ab}(t-s)\ =\ -\int
dx\,\big[(O^b\varrho_{\hspace{-0.01cm}_0})(x)\m\,L_x\,+\,
(j_{\hspace{-0.01cm}_0}\cdot\nabla O^b)(x)\big]\,P_{t-s}(x,dy)\m\,O^a(y)
\qqq
which, together with Eq.\,(\ref{prev}), implies the MFDT (\ref{mfd}).

\nsection{General fluctuation relations}

In \cite{Che}, two of us discussed arbitrary diffusion\ processes
in $d$ dimensions defined by the Stratonovich SDE
\begin{equation}
\dot{x}\ =\ u_t(x)\,+\,v_t(x)\,,  \label{sde}
\end{equation}
where $\,u_t(x)\,$ is a time-dependent deterministic vector field (a drift)
and $\,v_t(x)\,$ is a Gaussian random vector field with mean zero and
covariance
\begin{equation}
\big\langle v^{i}_t(x)\,v^{j}_s(y)\big\rangle\,=\,\delta
(t-s)\,D^{ij}_t(x,y)\,.  
\end{equation}
The Langevin equation (\ref{sys}) provides a special example of such an SDE.
\,For the processes solving Eq.\,(\ref{sde}), we showed, combining
the Girsanov and the Feynman-Kac formulae, a detailed
fluctuation relation (DFR)
\begin{equation}
\mu_{\hspace{-0.01cm}_0}(dx)
\,\,P_{\hspace{-0.05cm}_T}(x;dy,dW)\,\,\mathrm{e}^{-W}\,
=\,\mu_{\hspace{-0.01cm}_0}^{\prime
}(dy^{\ast })\,\,P_{\hspace{-0.05cm}_T}^{^{\prime }}(y^{\ast };
dx^{\ast },d(-W))\,,
\label{DFR}
\end{equation}
where
\hspace*{-0.3cm}{}

\begin{enumerate}
\item  $\,\mu_{\hspace{-0.01cm}_0}(dx)=\varrho_{\hspace{-0.01cm}_0}(x)
\,dx\,$ is the initial distribution of
the original (forward) process,

\item  $\,\mu_{\hspace{-0.01cm}_0}^{\prime }(dx)
=\varrho_{\hspace{-0.01cm}_0}^{\prime }(x)\,dx\,$ is the
initial distribution of the backward process obtained from the forward
process by applying a time inversion (see below),

\item  $\,P_{\hspace{-0.03cm}_T}(x;dy,dW)$ is the joint
probability distribution of the time
$\,T\,$ position $\,x_{\hspace{-0.03cm}_T}\,$ of the forward process
starting at time zero at $\,x\,$ and of a functional
$\,\CW_{\hspace{-0.03cm}_T}\,$ of the same process on the interval
$\,[0,T]\,$ (described later),

\item  $\,P_{\hspace{-0.04cm}_T}^{^{\prime }}(x;dy,dW)$ is the similar
joint probability
distribution for the backward process. 
\end{enumerate}

\noindent The key behind the DFR is the action of a time inversion on the
forward system. \,First, such an inversion acts on time and space by an
involution 
\qq
(t,x)\ \mapsto\ (T-t,x^{\ast })\,.
\label{stinv}
\qqq 
The above involution may be
extended to the action $\,\,x\mapsto\widetilde x\,\,$
on trajectories by the formula $\,\widetilde x_t=x^*_{T-t}\,$ and, further,
to the action on functionals of trajectories $\,\,\CF\mapsto\widetilde\CF\,$
\,by setting $\,\widetilde{\CF}[x]=\CF[\widetilde x]$. \,Second,
to recover a variety of fluctuation relations discussed in the literature
\cite {Kurchan,LebowSp,Crooks1,Crooks2,Jarz,SpS,CHCHJAR}, \,we allow for
a non-trivial behavior of the drift $\,u_t(x)\,$ under the time inversion,
dividing it into two parts, $\,u=u_++u_-$, \,with $\,u_{+}\,$ transforming
as a vector field under the space-time involution (\ref{stinv}) and 
$\,u_{-}\,$ as a pseudo-vector field, i.e. defining
\begin{eqnarray}
u'^{i}_{T-t,\pm}(x^{\ast })=\pm(\partial _{k}{x^*}%
^i)(x)\, \,u^k_{t,\pm}(x)\,,\qquad u^{\prime}=u^{\prime}_++\,u^{\prime}_-\,.
\end{eqnarray}
The random field $\,v_t(x)\,$ may be transformed with either rule. By
definition, the backward process satisfies then the Stratonovich SDE
\begin{equation}
\dot{x}\ =\ u^{\prime}_t(x)\,+\,v^{\prime}_t(x)
\end{equation}
and is, in general, different from the naive time inversion $\,\widetilde
x_t\,$ of the forward process. \,The functional
$\,\CW_{\hspace{-0.02cm}_T}\,$ involved
in the DFR depends explicitly on the densities
$\,\varrho_{\hspace{-0.01cm}_0}\,$ and
$\,\varrho_{\hspace{-0.01cm}_T}$, \,where the latter is defined
by the relation $\,\mu^{\prime}_{\hspace{-0.01cm}_0}(dx^*)=
\varrho_{\hspace{-0.01cm}_T}(x)\,dx\,$:
\begin{equation}
\CW_{\hspace{-0.02cm}_T}\ =\ -\Delta_{\hspace{-0.01cm}_T}
\ln\varrho\,+\,\int\limits_{0}^{T} \hspace{-0.1cm}
\CJ_t\,dt
\label{WT}
\end{equation}
with the notation $\,\Delta_{\hspace{-0.01cm}_T}\ln\varrho
\equiv\ln\varrho_{\hspace{-0.01cm}_T}(x_{\hspace{-0.02cm}_T})-
\ln\varrho_{\hspace{-0.01cm}_0}(x_{\hspace{-0.02cm}_0})$.
\,In the above formula,
\begin{equation}
\CJ_t\,=\,2\,\widehat{u}_{t,+}(x_t)\cdot d_t^{-1}(x_t)\big(\dot{x}_t
-u_{t,-}(x_t)\big)-(\nabla \cdot u_{t,-})(x_t)\,,
\end{equation}
where $\,d^{ij}_t(x)=D^{ij}_t(x,x)\,$ and $\,\widehat{u}_{t,+}^{i}(x)
=u_{t,+}^{i}(x)-\frac{1}{2} \partial
_{y^{j}}D^{ij}_t(x,y)|_{y=x}$. \,The time integral in Eq.\,(\ref{WT}) should
be taken in the Stratonovich sense. \,The functional
$\,\CW'_{\hspace{-0.02cm}_T}\,$
for the backward process is defined in the same way, \m setting
$\,\mu_{\hspace{-0.01cm}_0}(dx^*)
=\varrho'_T(x)dx$. \,One has the relation
\qq
\CW'_{\hspace{-0.02cm}_T}\,=\,-\widetilde\CW_{\hspace{-0.03cm}_T}\,,
\qqq
where the tilde denotes the involution of trajectory functionals
introduced before.
\vskip 0.1cm

The quantity $\,\CJ_t\,$ has the interpretation
of the rate of entropy production in the environment modeled by the thermal
noise. \,When the density $\,\varrho_{\hspace{-0.02cm}_T}\,$
coincides with the density obtained
from $\,\varrho_{\hspace{-0.01cm}_0}\,$ by the dynamical evolution
(\ref{coneq}), where now the current
\qq
j_t\,=\,(\hat u_t-\frac{_1}{^2}d_t\nabla)\varrho_t\qquad{\rm with}
\qquad \hat u\,=\,\hat u_+\,+\,u_-\,,
\label{curr}
\qqq
then the  first contribution $\,-\Delta_{\hspace{-0.01cm}_T}
\ln\varrho\,$ to $\,\CW_{\hspace{-0.02cm}_T}\,$
may be interpreted as the change in the instantaneous entropy of the process.
In this case, the functional $\,\CW_{\hspace{-0.02cm}_T}\,$ becomes
equal to the overall entropy production. Keep in mind that this is 
a fluctuating quantity which, in general, may take both positive 
and negative values.
\vskip 0.1cm

The DFR (\ref{DFR}) holds even if the measures $\,\mu_{\hspace{-0.01cm}_0}\,$
and $\,\mu_{\hspace{-0.01cm}_0}^{\prime }\,$ are not normalized,
or even not normalizable. For normalized initial measures, we denote by
\begin{eqnarray}
\big\langle\CF\big\rangle\,=\,\int\CF[x]\,\CM[dx]\qquad\ \mathrm{and}\ \qquad
\big\langle\CF\big\rangle^{\prime}\,=\,\int
\CF[x]\,\CM^{\prime}[dx]
\end{eqnarray}
the averages over the realizations of the forward and the backward process
$\,x_t,\ 0\leq t\leq T,\,$ with $\,x_{\hspace{-0.02cm}_0}\,$ distributed
according to the probability measure $\,\mu_{\hspace{-0.01cm}_0}\,$ and
$\,\mu^{\prime}_{\hspace{-0.01cm}_0}$, \,respectively.
$\,\CM[dx]\,$ and $\,\CM'[dx]\,$ stand for the corresponding measures over 
the space of trajectories. \,One of the  immediate consequences of
the DFR (\ref{DFR}) is the identity
\begin{equation}
\big\langle \mathrm{e}^{-\CW_{\hspace{-0.03cm}_T}}\big\rangle\ =\ 1  \label{JE}
\end{equation}
obtained by the integration of the both sides.
This is a generalization of the celebrated Jarzynski equality
\cite{Jarz1,Jarz}. The relation (\ref{JE}) implies the inequality
$\,\big\langle
\CW_{\hspace{-0.02cm}_T}\big\rangle\geq0\,$ that has
the form of the second law of
thermodynamics stating the positivity of the average entropy production. The
Jarzynski equality, however, provides also information about an
exponential suppression of the events with negative entropy production in
non-equilibrium systems for which $\,\big\langle\CW_{\hspace{-0.02cm}_T}
\big\rangle>0$.
\vskip 0.1cm

With a little more work \cite{Che} involving a multiple superposition of the
FDR (\ref{DFR}), the latter may be cast into the Crooks form \cite{Crooks2}
\begin{equation}
\big\langle\CF\,\mathrm{e}^{-\CW_{\hspace{-0.03cm}_T}}\big\rangle\ 
=\ \big\langle\widetilde{\CF}\m\big\rangle^{\prime}\m.
\label{Crooks}
\end{equation}
In terms of the trajectory measures
$\,\CM[dx]\,$ and $\,\CM^{\prime}[dx]$, \,this becomes the
identity
\begin{equation}
\widetilde{\CM}^{\prime }[dx]\ =\ \mathrm{e}^{-\CW_{\hspace{-0.03cm}_T}[x]}
\,\CM[dx]\,,
\label{RaNi}
\end{equation}
where $\,\widetilde{\CM}^{\prime }[dx]=\CM^{\prime }[d\widetilde{x}]$. \,
Eq.\,(\ref{RaNi}) permits to interpret the expectation of
$\,\CW_{\hspace{-0.02cm}_T}\,$ as the
relative entropy of the trajectory measures:
\begin{eqnarray}
\big\langle\CW_{\hspace{-0.02cm}_T}\big\rangle\ 
=\ \int\ln{\frac{\CM[dx]} {\widetilde
\CM^{\prime}[dx]}}\m\,\CM[dx]\ \equiv\ S(\CM|\widetilde\CM^{\prime})\,,
\end{eqnarray}
in line with the above entropic interpretation of the functional
$\,\CW_{\hspace{-0.02cm}_T}$.

\nsection{Fluctuation relations in Langevin dynamics}
\label{sec:6}

Let us specify the DFR (\ref{DFR}) to the case of Langevin dynamics (\ref
{LEq}) with, possibly, time-dependent Hamiltonian $\,H\,$ and
external force $\,G$. \,A canonical choice of the time inversion for
such a system takes
\qq
u_+\,=\,-\Gamma\nabla H\,,\qquad u_-\,=\,\Pi\nabla H\,+\,G
\qqq
and a linear involution $\,x^*=Rx\,$ such that
$\,R\Gamma R^t=\Gamma\,$ and $\,R\Pi R^t=-\Pi\,$ \cite{Che}. \,
For example, for the Langevin-Kramers dynamics in the phase-space,
the usual $\,R\,$ changes the sign of momenta. The backward dynamics
has now the same form as the forward one, with the time-reversed 
Hamiltonian $\,H'_t(x)=H_{T-t}(Rx)\,$ and the time-reversed external force 
$\,G'_t(x)=-RG_{T-t}(Rx)$. \,In this case, we may use the Gibbs
densities $\,\varrho_t(x)=Z_t^{-1}\ee^{-\beta H_t(x)}=\varrho'_{T-t}(Rx)\,$
at the initial and final times, with $\,Z_t\,$ standing for the partition 
function $\,\int\ee^{-\beta H_t(x)}dx\,$ if the integral is finite and
$\,Z_t=1\,$ otherwise. \,A straightforward calculation gives:
\begin{equation}
\CW_{\hspace{-0.02cm}_T}\,=\,\ln(Z_{\hspace{-0.02cm}_T}/
Z_{\hspace{-0.02cm}_0})\,
+\int\limits_{0}^{T}\big(\beta\m\partial_tH_t\,+\,\beta\m G_t
\cdot\nabla H_t\,- \,\nabla\cdot G_t\big)(x_t)\m\,dt\,.
\label{WTL}
\end{equation}
For normalizable Gibbs factors this is often called the ``dissipative work''.
$\,\CW'_{\hspace{-0.02cm}_T}\,$ is given by the same expression
with $\,H_t\,$ and $\,G_t\,$ replaced by $\,H'_t\,$ and $\,G'_t$.
\vskip 0.1cm

Another useful choice of the time inversion is based on the eventual knowledge
of the densities $\,\varrho_t\,$ corresponding to the conserved currents
with $\,\nabla\cdot j_t=0$. \,Note that such densities would be left invariant
by the evolution (\ref{coneq}) if the time-dependence of the Hamiltonian
and of the external force were frozen to the instantaneous values $\,H_t\,$
and $\,G_t$. \,One takes
\qq
u_+=\beta^{-1}\Gamma\nabla\ln\varrho\,,\qquad u_-\,=\,-\Gamma\nabla(H
+\beta^{-1}\ln\varrho)\,
+\,\Pi\nabla H\,+\,G\,.
\qqq
With the linear involution $x^*=Rx\,$ as above, the backward process has
\qq
u'_+\,=\,\beta^{-1}\Gamma\nabla\ln\varrho'\,,\qquad u'_-\,=\,
\Gamma\nabla(H'+\beta^{-1}\ln\varrho')\,+\,\Pi\nabla H'\,+\,G'\,,
\qqq
where $\,\varrho'_t(x)=\varrho_{T-t}(Rx)\,$ and $\,H'\,$ and $\,G'\,$ are
as before. \,The current corresponding
to the density $\,\varrho'_t\,$ satisfies
\qq
j'_t(x)\,=\,-R\,j_{\hspace{-0.02cm}_{T-t}}(Rx)\,.
\qqq
It is also conserved. Such a time inversion \,(for $\,R=1$) \,was
considered in \cite{HatSas} and, more explicitly, in \cite{CHCHJAR}.
\,In \cite{Che}, it was called the current reversal. The DFR (\ref{DFR})
holds now for
\qq
\CW_{\hspace{-0.02cm}_T}\,=\,
-\,\int\limits_0^T(\partial_t\ln\varrho_t)(x_t)\,dt
\label{WTC}
\qqq
and $\,\CW'_{\hspace{-0.02cm}_T}\,$ given by the same
expression with $\,\varrho_t\,$
replaced by $\,\varrho'_t$. \,The Jarzynski equality (\ref{JE}) for
this case (assuming that $\,\varrho_t\,$ are normalized) was first proven
by Hatano and Sasa in \cite{HatSas}. \,Note that if $\,G=0\,$ then the current
corresponding to the densities $\,\varrho_t=Z_t^{-1}\ee^{-\beta H_t}\,$
is conserved and with this choice of $\,\varrho_t\,$ the two time
inversions coincide. \,In particular, for $\,G=0\,$ and the time-independent
$\,H_t\equiv H$, \,the functional $\,\CW_{\hspace{-0.02cm}_T}\,$ identically
vanishes and the DFR reduces to the equality
\qq
\ee^{-\beta H(x)}\,dx\,\,P_{\hspace{-0.02cm}_T}(x,dy)\ =\ \ee^{-\beta H(y)}
\,dy\,\,P'_{T}(y^*,dx^*)
\qqq
which is a more global version of the detailed balance relation.
On the right hand side, one may replace $\,P'_T\,$ by 
$\,P_{\hspace{-0.02cm}_T}\,$ for $\,\Pi=0\,$ and $\,x^*\equiv x\,$
because in that case, the forward and backward processes have the
same distribution..

\subsection{Fluctuation relations for resurrecting processes}

Let us consider the process solving the SDE (\ref{SDEx}),
with $\,H_t(x)=ax^{k}+o(|x|^{k})\,$ at large $\,|x|\,$ with odd $\,k\geq 3\,$
and $\,a>0$. \,We admit a mild time-dependence of $\,H_t(x)\,$
disappearing when $\,x\to\pm\infty$. The corresponding process
still has a resurrecting version, as in the stationary case described
in Appendix A. \,Let us discuss first the canonical time
inversion with the trivial involution $\,x^*=x\,$ leading to the backward
process of the same type with $\,H_t(x)\,$ replaced by $\,H'_t(x)
=H_{T-t}(x)$. \,The definition of the functional
$\,\CW_{\hspace{-0.02cm}_T}\,$ for
the resurrecting process requires a little care in order to account
for the contributions from the jumps from $\,-\infty\,$ to $+\infty$.
\,This may be done by compactifying the line to a circle
writing $\,x=\cot{\theta}\,$ for $\,\theta\,$ modulo $\,\pi$. \,One has
\qq
\int\limits_0^T\CJ_t\,dt\,=\,-\beta\int\limits_0^T(\partial_xH_t)(x_t)\,
\dot{x}_t\,dt\,=\,
\int\limits_0^T\big(\beta\m k\m a\,\cot^{k-1}{\theta_t}\,+\,\dots\,\big)\,
\sin^{-2}{\theta_t}\m\,\dot{\theta_t}\,dt
\qqq
and the integral diverges to $\,+\infty\,$ whenever $\,\theta_t\,$ passes
from the negative to the positive values, \,i.e. whenever $\,x_t\,$
jumps from $\,-\infty\,$ to $\,+\infty$. \,Upon taking
$\,\varrho_t=\ee^{-\beta H_t}$, \,we infer that
\qq
\CW_{\hspace{-0.02cm}_T}\ =\ \begin{cases}\hbox to 3.1cm{$\,\int\limits_0^T
(\partial_tH_t)(x_t)\,dt$\hfill}
{\rm if}\quad x_t\ \,{\rm has\ no\ rebirths\ for}
\ \,0< t< T\,,\cr
\hbox to 3.1cm{$\,+\infty$\hfill}{\rm otherwise\,.}\end{cases}
\qqq
and similarly for $\,\CW'_{\hspace{-0.02cm}_T}$. \,As a result,
the contributions of the
rebirths to the DFR (\ref{DFR}) trivially decouple reducing the latter
to the identity
\qq
\ee^{-H_0(x)}dx\,\,P^0_{\hspace{-0.03cm}_T}(x;dy,dW)
\,\,\ee^{-W}\ =\,\ \ee^{-H_{\hspace{-0.03cm}_T}(y)}dy\,\,
P'^0_{\hspace{-0.03cm}_T}(x;dy,d(-W))
\label{DFR0}
\qqq
between the joint distributions of the endpoints and of the functional
$\,\,\CW_{\hspace{-0.02cm}_T}=\int_0^T(\partial_tH_t)(x_t)\,\,$ (usually
called the ``Jarzynski work''), \,or of its counterpart 
$\,\CW'_{\hspace{-0.02cm}_T}$, \m\,in the processes without rebirths. 
\m In the stationary case,
\m Eq.\,(\ref{DFR0}) reduces to the detailed balance relation
\qq
\ee^{-\beta H(x)}dx\,\,P^0_t(x,dy)\ =\ \ee^{-\beta H(y)}dy\,\,P^0_t(y,dx)\,.
\label{DB0}
\qqq
which assures upon integration over $\,x\,$ that the process without
rebirths preserves the infinite measure $\,\ee^{-\beta H(x)}dx$.
\vskip 0.1cm

On the other hand, one could use for the same SDE with the resurrecting
solution the current reversal based on the splitting
\qq
u_{t,+}\,=\,\beta^{-1}\partial_x\ln\varrho_{\hspace{-0.01cm}_{H_t}}\,,
\qquad u_{t,-}\,=\,-\partial_x(H_t+\beta^{-1}
\ln{\varrho_{\hspace{-0.01cm}_{H_t}}})
\qqq
of the drift $\,-\partial_xH_t$, \,with the density
$\,\varrho_{\hspace{-0.01cm}_{H_t}}\,$
given by Eq.\,(\ref{C1}). \,The use of the involution $\,x^*=-x\,$
leads then to the backward process solving the SDE (\ref{SDEx}) with
the Hamiltonian $\,H_t(x)\,$ replaced by
\qq
H'_t(x)\,=\,H_{\hspace{-0.01cm}_{T-t}}\hspace{-0.01cm}(-x)\,
-\,2\beta^{-1}\ln\Big(\int\limits_{-\infty}^{-x}
\ee^{\m\beta H_{\hspace{-0.01cm}_{T-t}}(y)}dy\Big)\,.
\qqq
From the estimate (\ref{estim}) in Appendix A, one infers that
$\,H'_t(x)=ax^{k}+o(x^{k-1})\,$ for large $|x|\,$
(we have used the non-trivial spatial involution to keep $\,a\,$
positive). \,Hence $\,H'_t\,$ is of the same type as the Hamiltonian 
$\,H_t\,$ for the forward process. In this case, the functionals 
$\,\CW_{\hspace{-0.03cm}_T}\,$ and $\,\CW'_{\hspace{-0.02cm}_T}\,$
are given by Eq.\,(\ref{WTC}) with $\,\varrho_t(x)\,$ replaced by
$\,\varrho_{\hspace{-0.02cm}_{H_t}}\hspace{-0.03cm}(x)\,$ and
$\,\varrho_{\hspace{-0.02cm}_{H_{T-t}}}\hspace{-0.03cm}(-x)
=\varrho_{\hspace{-0.02cm}_{H'_t}}\hspace{-0.02cm}(x)$,\,
respectively, \,with no extra contributions from the rebirths.
\,In the stationary case, one has $\,\CW_{\hspace{-0.02cm}_T}
=0=\CW'_{\hspace{-0.02cm}_T}\,$ and the DFR
(\ref{DFR}) reduces to the modified detailed balance relation
\qq
\varrho_{\hspace{-0.02cm}_H}\hspace{-0.01cm}(x)\,dx\,\,
P_{\hspace{-0.03cm}_T}(x,dy)\,
=\,\varrho_{\hspace{-0.02cm}_H}\hspace{-0.01cm}(y)\,dy\,\,
P'_{\hspace{-0.02cm}_T}(-y,d(-x))\,.
\qqq
The latter links the transition probabilities of the resurrecting forward
and backward processes and assures upon integration over $\,x\,$
or $\,y\,$ that those processes preserve the probability measures
$\,\varrho_{\hspace{-0.02cm}_H}\hspace{-0.03cm}(x)\,dx\,$ and
$\,\varrho_{\hspace{-0.02cm}_H}\hspace{-0.03cm}(-x)\,dx$,
\,respectively.

\subsection{Fluctuation relations for forced diffusions on circle}

Consider the process satisfying the SDE (\ref{1dLan}) with periodic
Hamiltonian $\,H(x)=H(x+2\pi)\,$ and external force $\,G(x)=G(x+2\pi)$,
\,both possibly time-dependent. The use of the canonical time inversion
with the trivial involution $\,x^*=x\,$ leads to the backward process
of the same type with $\,H'_t=H'_{T-t}\,$ and $\,G'_t=-G'_{T-t}$.
\,The functionals $\,\CW_{\hspace{-0.03cm}_T}\,$
and $\,\CW'_T\,$ are given here by
the formula (\ref{WTL}).
\vskip 0.1cm

On the other hand, the use of the current reversal with
the densities $\,\varrho_{\hspace{-0.02cm}_{H_t}}$ of Eq.\,(\ref{C2})
and the trivial inversion $\,x^*=x\,$ leads to the backward
process satisfying the same SDE with $\,H_t\,$ replaced by
\qq
H'_t(x)\,=\,-H_{\hspace{-0.02cm}_{T-t}}\hspace{-0.01cm}(x)\,
-2\beta^{-1}\ln\varrho_{_{H_{\hspace{-0.02cm}_{T-t}}}}
\qqq 
and $\,G_t\,$ by $\,G'_t\,$ as above. The functionals
$\,\CW_{\hspace{-0.02cm}_T}\,$ and $\,\CW'_{\hspace{-0.02cm}_T}\,$
are given now by Eq.\,(\ref{WTC}) with
$\,\varrho_t(x)\,$ equal to $\,\varrho_{\hspace{-0.02cm}_{H_t}}
\hspace{-0.04cm}(x)\,$ and $\,\varrho_{H'_t}(x)=
\varrho_{\hspace{-0.02cm}_{H_{T-t}}}
\hspace{-0.04cm}(x)$, \,respectively. They vanish in the stationary
case when the DFR (\ref{DFR}) reduces again the modified detailed
balance relation
\qq
\varrho_{\hspace{-0.02cm}_H}\hspace{-0.01cm}(x)\,dx\,\,
P_{\hspace{-0.03cm}_T}(x,dy)\,
=\,\varrho_{\hspace{-0.02cm}_H}\hspace{-0.01cm}(y)\,dy\,\,
P'_{\hspace{-0.02cm}_T}(y,dx)\,.
\qqq
Note that, in general, to obtain the DFR for the probability distributions 
on the circle, one should sum both sides of the relation (\ref{DFR})
pertaining to the motion on the line, over the shifts of $\,x\,$ or 
$\,y\,$ by $\,2\pi n\,$ with integer $\,n\,$ (both summations amount 
to the same). To get the Jarzynski equality, one has to integrate 
the relation obtained 
this way over $\,x\,$ and $\,y\,$ from 0 to $2\pi$. \,Another simple
remark is that for the SDE (\ref{1dLan}) on the circle, one may always
assume that the external force $\,G\,$ is constant by changing
$\,G(x)\,$ to $\,\overline{G}=\frac{1}{2\pi}\int_0^{2\pi} G(x)\,dx\,$
and by subtracting $\m\,\int_0^x(G(y)-\overline{G})\m dy\m\,$
from $\,H(x)$. \,Such a change does not affect the DFR (\ref{DFR})
obtained by the current reversal that uses only the invariant
densities and it modifies in a simple way the DFR obtained from 
the canonical time inversion because in the latter we used the Gibbs 
measures for the initial and final distributions.

\nsection{Fluctuation relations close to NESS}

As promised, we shall show here that the MFDT (\ref{mfd}), proven
directly in Sec.\,4.1, may be also derived by reducing the Crooks DFR 
for the current reversal, and, in a special case, the 
Jarzynski-Hatano-Sasa equality, to the situations close
to NESS.

\subsection{Reduction of Crooks' DFR to MFDT}

Let us consider the DFR (\ref{Crooks}) for the Langevin dynamics
(\ref{sys}) with $\,\CF=O^a_t\,$ and $\,0<t<T$,
\,the backward dynamics determined by
the current reversal, and, for simplicity, the trivial 
space involution $\,x^*\equiv x$, \,see Sec.\,\ref{sec:6}. 
\,It reads:
\qq
\big\langle O^a_t\,\ee^{-W_{\hspace{-0.03cm}_T}}\big\rangle\,
=\,\big\langle O^a_{T-t}\big\rangle'\m.
\label{Crtop}
\qqq
We shall assume
that the time-dependence of the Hamiltonian is given by
Eq.\,(\ref{hpert}) with $\,h_{a,0}=0=h_{a,T}$, \,and that the external
force $\,G\,$ is time-independent. Let, as above,
$\,\varrho_{\hspace{-0.01cm}_0}(x)\,dx\,$
denote the invariant probability measure of the unperturbed process
(assumed to exist). By $\,\varrho_t$, we shall denote now
the normalized densities whose current $\,j_t$, \,given by
Eq.\,(\ref{curr0}), is conserved, \,i.e. such that
\qq
L^\dagger\varrho_t\,-\,\sum\limits_a
h_{t,a}\nabla\cdot\big((\Gamma-\Pi)(\nabla O^a)\,
\varrho_t\big)\,
=\,0\,.
\label{11}
\qqq
Expanding
\qq
\varrho_t\,=\,\varrho_{\hspace{-0.01cm}_0}\,
+\,\sum\limits_ah_{t,a}\m\varrho^{a}_1\ +\ \CO(h^2)\,,
\qqq
we obtain from Eq.\,(\ref{11}) the relation
\qq
L^\dagger\varrho^{a}_1\,=\,\nabla\cdot\big((\Gamma-\Pi)(\nabla O^a)
\,\varrho_{\hspace{-0.01cm}_0}\big)
\qqq
whose unique solution
\qq
\varrho^{a}_1\,=\,(L^\dagger)^{-1}\big[
\nabla\cdot\big((\Gamma-\Pi)(\nabla O^a)\,\varrho_{\hspace{-0.01cm}_0}
\big)\big]
\label{unisol}
\qqq
is chosen by imposing the orthogonality of $\,\varrho^{a}_1\,$
to the constant mode, required by the normalization of $\,\varrho_t$. \,
For the current reversal, the functional $\,\CW_{\hspace{-0.03cm}_T}\,$
is given by Eq.\,(\ref{WTC}) so that
\qq
&&\CW_{\hspace{-0.02cm}_T}\ =\ -\sum\limits_a
\int\limits_0^T\big(\partial_t{h}_{t,a}\big)
\,\big(\varrho_{\hspace{-0.01cm}_0}^{-1}\varrho^a_1
\big)_t\m\,dt\ +\ \CO(h^2)\cr
&&=\ \sum\limits_a\int\limits_0^Th_{t,a}\m\,\partial_t
\big\{\varrho_{\hspace{-0.01cm}_0}^{-1}(L^\dagger)^{-1}\big[
\nabla\cdot\big((\Gamma-\Pi)(\nabla O^a)\,\varrho_{\hspace{-0.01cm}_0}
\big)\big]\big\}_t\,
\m\,dt\ +\ \CO(h^2)\,,
\qqq
where we have integrated ones by parts over $\,t$.
\,The application of the operator $\,\frac{\delta}{\delta h_{b,s}}|_{h=0}\,$
for $\,0<s<t\,$ to the both sides of Eq.\,(\ref{Crtop}) gives the identity
\qq
\CR^{ab}(t-s)\,-\,\big\langle
O^a_t\,\partial_s\big\{\varrho_{\hspace{-0.01cm}_0}^{-1}
(L^\dagger)^{-1}\big[
\nabla\cdot\big((\Gamma-\Pi)(\nabla O^a)\,
\varrho_{\hspace{-0.01cm}_0}\big)\big]\big\}_s
\big\rangle_{\hspace{-0.03cm}0}\ =\ 0\,.
\label{relu}
\qqq
We have used the fact that, by causality, the right hand
side of Eq.\,(\ref{Crtop}) does not give the contribution
because the perturbation is concentrated around time $\,(T-s)>(T-t)$.
\,From the relation (\ref{relu}), it follows that
\qq
&&\CR^{ab}(t-s)\ =\
\big\langle
O^a_t\,\partial_s\big\{\varrho_{\hspace{-0.01cm}_0}^{-1}
(L^\dagger)^{-1}\big[
\nabla\cdot\big((\Gamma-\Pi)(\nabla O^a)\,
\varrho_{\hspace{-0.01cm}_0}\big)\big]\big\}_s
\big\rangle_{\hspace{-0.03cm}0}\cr
&&=\ \partial_s\int dx\,\big\{(L^\dagger)^{-1}\big[
\nabla\cdot\big((\Gamma-\Pi)(\nabla O^a)\,
\varrho_{\hspace{-0.01cm}_0}\big)\big]\big\}(x)
\m\,P_{t-s}(x,dy)\m\,O^a(y)\cr
&&=\ -\int dx\,\big\{(L^\dagger)^{-1}\big[
\nabla\cdot\big((\Gamma-\Pi)(\nabla O^a)\,
\varrho_{\hspace{-0.01cm}_0}\big)\big]\big\}(x)
\m\,L_xP_{t-s}(x,dy)\m\,O^a(y)\cr
&&=\ -\int dx\,\big[\nabla\cdot\big((\Gamma-\Pi)(\nabla O^a)\,
\varrho_{\hspace{-0.01cm}_0}\big)\big](x)\m\,P_{t-s}(x,dy)\m\,O^a(y)
\qqq
which is Eq.\,(\ref{resp1}) above. The rest of the proof of the identity
(\ref{mfd}) goes as before.

\subsection{Jarzynski-Hatano-Sasa equality and MFDT}

The standard FDT around the equilibrium Langevin dynamics (\ref{sys})
without the external force may be obtained by expanding the
Jarzynski equality (\ref{JE}) for the Hamiltonian (\ref{hpert})
up to the second order in $\,h$, \,see \cite{Che}. The Jarzynski
equality may then be viewed as an extension of the FDT to the case
of Hamiltonians with arbitrary time dependence driving the system
far from equilibrium. The natural question is whether this picture
may be generalized to the case of the modified FDT (\ref{mfd})
holding around NESS. We shall show here that the answer is 
a qualified yes.
\vskip 0.1cm

Let us expand to the second order in $\,h\,$ the Hatano-Sasa
version of the Jarzynski equality (\ref{JE}) obtained from the Croocks DFR
(\ref{Crtop}) for the current reversal by replacing $\,O^a\,$ by $\,1$.
\,We shall need to know the form of the densities $\,\varrho_t\,$
with conserved current, i.e. satisfying Eq.\,(\ref{11}), to
the second order in $\,h$. \,One has
\qq
\varrho_t\,=\,\varrho_{\hspace{-0.01cm}_0}\,+\,\sum\limits_ah_{t,a}
\m\varrho^{a}_1\,
+\,\sum\limits_{a,b}h_{t,a}\m h_{t,b}\m\varrho^{ab}_2\ +\ \CO(h^3)
\qqq
with $\,\varrho_{1a}\,$ as before, see Eq.\,(\ref{unisol}), and
\qq
\varrho^{ab}_2\,=\,(L^+)^{-1}\big[\nabla\cdot\big((\Gamma-\Pi)(\nabla O^a)
\,\varrho^b_1\big)\big]\m.
\qqq
Expanding, in turn, the functional $\,\CW_{\hspace{-0.03cm}_T}\,$ given
by Eq.\,(\ref{WTC}) to the second order, we obtain
\qq
\CW_{\hspace{-0.02cm}_T}\ =\ -\sum\limits_a
\int\limits_0^T\big(\partial_t h_{t,a}\big)\,
\big(\varrho_{\hspace{-0.01cm}_0}^{-1}\varrho^a_1\big)_t\m\,dt\,
-\sum\limits_{a,b}\int\limits_0^T\big(\partial_t(h_{t,a}\m h_{t,b})\big)\,
\big(\varrho_{\hspace{-0.01cm}_0}^{-1}\varrho^{ab}_2-\frac{_1}{^2}
\varrho_{\hspace{-0.01cm}_0}^{-2}\varrho^a_1
\varrho^b_1\big)_t\m\,dt\ +\ \CO(h^3)\,,\quad
\qqq
and further,
\qq
\Big\langle\ee^{-\CW_{\hspace{-0.03cm}_T}}\Big\rangle\,-\,1&=&
\sum\limits_a\int\limits_0^T\big(\partial_t h_{t,a}\big)\,
\Big\langle\big(\varrho_{\hspace{-0.01cm}_0}^{-1}
\varrho^a_1\big)_t\Big\rangle\,\,dt
\,+\sum\limits_{a,b}\int\limits_0^T\big(\partial_t(h_{t,a}\m h_{t,b})\big)\,
\Big\langle\big(\varrho_{\hspace{-0.01cm}_0}^{-1}
\varrho^{ab}_2-\frac{_1}{^2}\varrho_{\hspace{-0.01cm}_0}^{-2}
\varrho^a_1\varrho^b_1\big)_t\Big\rangle_{\hspace{-0.05cm}0}\,dt\cr
&+&\frac{_1}{^2}\sum\limits_{a,b}\int\limits_0^T\int\limits_0^T
(\partial_t h_{t,a})\,(\partial_s h_{s,b})
\,\,\Big\langle\big(\varrho_{\hspace{-0.01cm}_0}^{-1}\varrho^a_1\big)_t\,
\big(\varrho_{\hspace{-0.01cm}_0}^{-1}\varrho^b_1\big)_s
\Big\rangle_{\hspace{-0.05cm}0}
\,dt\,ds\ \,+\,\ \CO(h^3)\,.\ \quad
\qqq
The second term on the right hand side integrates to zero
because of the stationarity of the unperturbed expectation and
the boundary conditions $\,h_{0,a}=0=h_{T,a}$.
\,Expanding the remaining perturbed expectation in the first
term on the right hand side, we infer that
\qq
\Big\langle\ee^{-\CW_{\hspace{-0.03cm}_T}}\Big\rangle\,-\,1&=&
\sum\limits_{a,b}\int\limits_0^T\int\limits_0^T(\partial_th_{t,a})\,
h_{s,b}\,\Big\langle\big(\varrho_{\hspace{-0.01cm}_0}^{-1}
\varrho^a_1\big)_t\m\,R^b_s\Big\rangle_{\hspace{-0.05cm}0}\,\,dt\,ds\cr
&+&\frac{_1}{^2}\sum\limits_{a,b}\int\limits_0^T\int\limits_0^T
(\partial_th_{t,a})\,(\partial_s h_{s,b})
\,\,\Big\langle\big(\varrho_{\hspace{-0.01cm}_0}^{-1}\varrho^a_1\big)_t\,
\big(\varrho_{\hspace{-0.01cm}_0}^{-1}
\varrho^b_1\big)_s\Big\rangle_{\hspace{-0.05cm}0}
\,dt\,ds\ \,+\,\ \CO(h^3)\ =\ 0\,,\qquad
\qqq
where the last equality follows from the (generalized) Jarzynski
equality (\ref{JE}). \,The integration by parts and the causality 
permit to conclude that for $\,t>s$,
\qq
\partial_t\Big\langle (\varrho_{\hspace{-0.01cm}_0}^{-1}\varrho_1^a)_t\,R^b_s
\Big\rangle_{\hspace{-0.05cm}0}\ =\ \partial_t\partial_s
\Big\langle(\varrho_{\hspace{-0.01cm}_0}^{-1}\varrho_1^a)_t\,
(\varrho_{\hspace{-0.01cm}_0}^{-1}\varrho_1^b)_s
\Big\rangle_{\hspace{-0.05cm}0}\,,
\qqq
or, \,integrating ones over $\,t$, \,that
\qq
\Big\langle (\varrho_{\hspace{-0.01cm}_0}^{-1}\varrho_1^a)_t\,R^b_s
\Big\rangle_{\hspace{-0.05cm}0}\ =\ \partial_s
\Big\langle(\varrho_{\hspace{-0.01cm}_0}^{-1}\varrho_1^a)_t
\,(\varrho_{\hspace{-0.01cm}_0}^{-1}\varrho_1^b)_s
\Big\rangle_{\hspace{-0.05cm}0}
\label{aftin}
\qqq
(we used the fact that both sides vanish for $\,t=s$).
\,Note that Eq.\,(\ref{aftin}) stays true if we add to $\,\varrho^a_1\,$ any
multiple of $\,\varrho_0$, \,so that we may drop the normalization
condition $\,\int\varrho^a_1(x)\,dx=0$. \,From Eqs.\,(\ref{unisol})
and (\ref{fstal}), it follows
then that we may take
\qq
\varrho_{\hspace{-0.01cm}_0}^{-1}\varrho_1^a\
=\ \beta\m(O^a-\widehat{O}^a)\ \equiv\
\beta\m A^a\,,
\qqq
where the dressed observable
\qq
\widehat{O}^a\,=\,-\varrho_{\hspace{-0.01cm}_0}^{-1}(L^\dagger)^{-1}
\big[j_0\cdot\nabla O^a\big]\,.
\label{hat}
\qqq
The identity (\ref{aftin}) becomes now the relation
\qq
\beta^{-1}\big\langle A^a_t\,R^b_s\big\rangle_{\hspace{-0.03cm}0}&=&
\partial_s\big\langle A^a_t\,A^b_s\big\rangle_{\hspace{-0.03cm}0}\ =\
\partial_s\big\langle A^a_t\,O^b_s\big\rangle_{\hspace{-0.03cm}0}\,-\,
\partial_s\int dx\,\,(\widehat{O}^b\varrho_{\hspace{-0.01cm}_0})(x)
\m\,P_{t-s}(x,dy)\m\,A^a(y)\cr
&=&\,\partial_s\big\langle A^a_t\,O^b_s\big\rangle_{\hspace{-0.03cm}0}\,
+\,\int dx\,\,(\widehat{O}^b\varrho_{\hspace{-0.01cm}_0})(x)
\,\,L_xP_{t-s}(x,dy)\m\,A^a(y)\,.
\qqq
Integrating by parts in the last term and using the definition
(\ref{hat}), we finally obtain the identity
\qq
\beta^{-1}\big\langle A^a_t\,R^b_s\big\rangle_{\hspace{-0.03cm}0}&=&
\partial_s\big\langle
A^a_t\,O^b_s\big\rangle_{\hspace{-0.03cm}0}\ -\
\int dx\,\,
(j_{\hspace{-0.01cm}_0}\cdot\nabla O^b)(x)\,\,
P_{t-s}(x,dy)\m\,A^a(y)\cr
&=&
\partial_s\big\langle
A^a_t\,O^b_s\big\rangle_{\hspace{-0.03cm}0}\,-\,
\big\langle A^a_t\,B^b_s\big\rangle_{\hspace{-0.03cm}0}
\qqq
which is the MFDT (\ref{mfd}) with the observable $\,O^a\,$
replaced by $\,A^a$. \,It is a consequence of the
identity (\ref{mfd}) but, in general, it does not seem to be
equivalent to it, except for the equilibrium case with vanishing
external force and $\,\varrho_{\hspace{-0.01cm}_0}
=Z^{-1}\ee^{-\beta H}\,$ when
$\m\,j_{\hspace{-0.01cm}_0}=0\,\,$
and $\,\,A^a=O^a$.
\vskip 0.1cm

In the special case of the one-dimensional NESS
with constant current $\,j_{\hspace{-0.01cm}_0}\,$ described above,
\,the dressing (\ref{hat}) of the observables
coincides with the one given by Eqs.\,(\ref{hat12})
for the types 1 and 2, respectively. This may be
easily seen by checking that for the latter,
\qq
L^\dagger\varrho_{\hspace{-0.01cm}_0}\widehat{V}\,
=\,-j_{\hspace{-0.01cm}_0}\,\partial_xV
\qqq
with $\,\varrho_{\hspace{-0.01cm}_0}=\varrho_{\hspace{-0.01cm}_H}\,$
given by Eqs.\,(\ref{C1}) and (\ref{C2}).

\nsection{Conclusions}

We have discussed different fluctuation relations for the Langevin dynamics.
Those included an extension of the fluctuation-dissipation theorem (FDT)
(\ref{EFDT}), one of the most important relations of the (close to) 
equilibrium statistical mechanics, to the case of non-equilibrium steady 
states (NESS) of Langevin processes. The modified fluctuation-dissipation
theorem (MFDT) (\ref{mfd}) that holds around NESS has a new term containing 
the probability current but in the Lagrangian frame moving with the mean 
local velocity determined by the current, it takes the form (\ref{extend}) 
similar to that of the equilibrium FDT. We also pointed out that, similarly 
to the equilibrium FDT, the MFDT may be viewed as a limiting case of more 
general fluctuation relations that are valid arbitrarily far from the 
stationary situation, namely of the Crooks detailed fluctuation relation 
(\ref{Crooks}) for the backward dynamics with inverted probability current 
or of the Jarzynski-Hatano-Sasa equality (\ref{JE}). The general discussion 
was illustrated on two examples of one-dimensional systems with explicit 
non-equilibrium invariant measures.

\appendix

\newcommand{\appsection}[1]{\let\oldthesection\thesection
  \renewcommand{\thesection}{Appendix \oldthesection}
  \section{#1}\let\thesection\oldthesection\setcounter{equation}{0}}

\appsection{}

We return here to the case of the stationary SDE (\ref{SDEx}).
The transition probabilities $\,P_{t}(x,dy)\,$ for the diffusion
process solving this equation are given by the kernels of the exponential
of the generator $\,L=\beta^{-1}\partial_x^2-(\partial_xH)\partial_x\,$
of the process:
\qq
\int P_{t}(x,dy)\,f(y)\,=\,\big(\ee^{\m tL}f\big)(x)\,,
\qqq
see Eq.\,(\ref{prev}). One has the following relation
\qq
L\,=\,-\beta^{-1}\ee^{\frac{1}{2}\beta H}Q^\dagger Q\,
\ee^{-\frac{1}{2}\beta H}
\qqq
for $\,Q=\partial_x+\frac{1}{2}\beta(\partial_xH)$, $\,
Q^\dagger=-\partial_x+\frac{1}{2}\beta(\partial_xH)$.
\,The Fokker-Planck operator
\qq
\beta^{-1}Q^\dagger Q\,=\,-\beta^{-1}\partial_x^2
+\frac{_1}{^2}(\partial_x^2H)
+\frac{_1}{^4}\beta(\partial_xH)^2
\qqq
is a positive self-adjoint Hamiltonian of a super-symmetric quantum
mechanics \cite{Witten} so that the transition probabilities may be
defined by the relation
\qq
P_{t}(x,dy)\ =\ \ee^{\m\frac{1}{2}\beta H(x)}\,\,\ee^{-t\beta^{-1}
Q^\dagger Q}(x,dy)\,\,\ee^{-\frac{1}{2}\beta H(y)}
\label{trpr}
\qqq
If the Gibbs density is normalizable with $\,Z=\int\ee^{-H(x)}dx<\infty\,$
then $\,\psi_{\hspace{-0.01cm}_0}(x)
=Z^{-1/2}\ee^{-\frac{1}{2}\beta H(x)}\,$ provides the
zero-energy groundstate of the Fokker-Planck Hamiltonian. Such a groundstate
is supersymmetric:
$\,Q\psi_{\hspace{-0.01cm}_0}=0$. 
\,For the Hamiltonian $\,H(x)=ax^{k}+o(|x|^{k})\,$
at large $\,|x|\,$ with either even $\,k\geq2\,$ and $\,a<0\,$ or odd
$\,k\geq 3$, \,however, the groundstate $\,\psi_{\hspace{-0.01cm}_0}\,$ 
of $\,\beta^{-1}
Q^\dagger Q\,$ is not given by $\,\ee^{-\frac{1}{2}\beta H(x)}\,$ but
has positive energy $\,E_{\hspace{-0.01cm}_0}>0\,$ and breaks the 
supersymmetry: $\,Q\psi_{\hspace{-0.01cm}_0}\not=0$.
\,In these cases, the transition probabilities (\ref{trpr}) are not
normalized for $\,t>0\,$ with
\qq
1\ >\ \int P_t(x,dy)\ \ \ \mathop{\sim}\limits_{t\ {\rm large}}
\ \ \ \ee^{-E_{\hspace{-0.01cm}_0}t}\,.
\qqq
The defect $\,\big(1-\int P_t(x,dy)\big)\,$ gives the probability
that the diffusion
process $\,x_t\,$ solving the SDE (\ref{SDEx}) and starting at time zero
at $\,x\,$ escapes by time $\,t\,$ to $\,\pm\infty$.
\,Writing $\,P_t(x,dy)\equiv P_t(x,y)\,dy$, \,the probability that the
escape happens between times $\,s\,$ and $\,s+ds\,$ may be expressed as
\qq
\Big(-\frac{_d}{^{ds}}\int P_s(x,dy)\Big)\,ds&=&\Big(-\int
\partial_y\big[\big(\beta^{-1}\partial_y+(\partial_yH)(y)\big)\,P_s(x,y)\big]\,dy
\Big)\,ds\cr
&=&\Big(-\big(\beta^{-1}\partial_y+(\partial_yH)(y)\big)
\,P_s(x,y)\big|_{y=-\infty}^{y=\infty}\Big)\,ds
\qqq
with the $\,y=\pm\infty\,$ terms determining the rates of escape 
to $\pm\infty$, \,respectively.
\vskip 0.1cm

Let us concentrate on the case with $\m\,H(x)=a x^{k}+o(|x|^{k})\,\,$
for odd $\,k\geq 3\,$ and $\,a>0$, \,denoting
the transition probabilities of Eq.\,(\ref{trpr}) by $\,P^0_t(x,dy)$.
\,Although they are not given by a closed analytic expression, their time
integral, equal to the Green kernel of $\,L$, \,is:
\qq
\int\limits_0^\infty ds\,\,P^0_s(x,dy)\,=\,(-L)^{-1}(x,dy)\,=\,
\beta\Big(\hspace{-0.1cm}\int\limits_{-\infty}^{{\rm min}(x,y)}\hspace{-0.3cm}
\ee^{\m\beta H(z)}dz\Big)\,\ee^{-\beta H(y)}dy\,.
\label{Green}
\qqq
From the last formula, we infer that
\qq
\lim\limits_{y\to\pm\infty}\,\int\limits_0^\infty ds\,\,(\beta^{-1}
\partial_y+(\partial_yH)(y))\m P^0_s(x,dy)\ =\ \delta_{1,\mp 1}
\qqq
so that the process $\,x_t\,$ escapes here only to $\,-\infty\,$
(this is already true if we ignore the noise in Eq.\,(\ref{SDEx})).
On the other hand, since the limit of the right hand side of 
\m Eq.\,(\ref{Green}) when $\,x\to+\infty\,$ exists, it follows that 
the transition probabilities from $\,x=+\infty$,
\qq
P^0_t(\infty,dy)\,=\,\lim\limits_{x\to+\infty}P^0_t(x,dy)
\qqq
are finite, non-zero measures. \,One may then define a
resurrecting version of the process $\,x_t\,$ solving the SDE
(\ref{SDEx}). \,The trajectories of such a process,
after almost surely reaching $\,-\infty\,$ reappear immediately
at $+\infty$. \,The resurrecting process is Markov and its
transition probabilities are
\qq
P_t(x,dy)\,=\,\sum\limits_{n=0}^\infty P^n_t(x,dy)\,,
\qqq
where $\,P^n_t(x,dy)\,$ are the transition
probabilities  with exactly $\,n\,$ jumps from $\,-\infty\,$ to $\,+\infty$.
\,They are given by the recursion relation:
\qq
P^{n+1}_t(x,dy)\ =\ -\int\limits_0^t ds\,\Big(\frac{_d}{^{ds}}\int P^0_s(x,dy)
\Big)\,P^n_{t-s}(\infty,dy)\,.
\label{rec}
\qqq
In terms of the Laplace transforms
\qq
\hat P^n_\omega(x,dy)\,=\,\int\limits_0^\infty\ee^{-t\omega}\, P^n_t(x,dy)\,,
\qqq
the recursion (\ref{rec}) becomes the equality
\qq
\hat P^{n+1}_\omega(x,dy)\ =\ \Big(1-\omega\int\hat P^0_\omega(x,dz)\Big)
\,\hat P^n_\omega(\infty,dy)
\qqq
which may be easily solved by iteration:
\qq
\hat P^n_\omega(x,dy)\,=\,\Big(1-\omega\int\hat P^0_\omega(x,dz)
\Big)^{\hspace{-0.03cm}n}
\,\hat P^0_\omega(\infty,dy)
\qqq
for $\,n\geq 1$. \,Re-summing the geometric progression, one obtains
for the Laplace transform of the transition probabilities of
the resurrecting process the expression
\qq
\hat P_\omega(x,dy)\,=\,\hat P^0_\omega(x,dy)\,+\,\frac{\Big(
1-\omega\int\hat P^0_\omega(x,dz)\Big)\hat P^0_\omega(\infty,dy)}{\omega
\int\hat P^0_\omega(\infty,dz)}\,.
\qqq
Note that $\,\hat P^0_\omega(x,dy)\,$ is analytic in $\,\omega\,$
for $\,{\rm Re}\m\omega>-E_{\hspace{-0.01cm}_0}\,$ 
but $\,\hat P_\omega(x,dy)\,$ has a pole
at $\,\omega=0\,$ with the residue
\qq
\mu(dy)\ =\ \frac{\hat P^0_0(\infty,dy)}{\int\hat P^0_0(\infty,dz)}\ =\
\frac{1}{Z}\,\Big(\int\limits_{-\infty}^y\ee^{\m\beta H(z)}\,dz\Big)
\,\ee^{-\beta H(y)}\,dy\,,
\qqq
where $\,Z\,$ is the normalization constant. This is the invariant
probability measure (\ref{C1}) of the resurrecting process.
\vskip 0.1cm

Let us finish by estimating the behavior of the density of the
invariant measure $\,\mu(dy)\,$ when
$\,|y|\to\infty$. \,We shall show that
\qq
\int\limits_{-\infty}^y\ee^{\m\beta(H(z)-H(y))}\,dz\ =\
\frac{1}{a\beta k\m y^{k-1}}\ +\ o(y^{-k+1})\,.
\label{estim}
\qqq
To this end, we rewrite the latter integral as
\qq
\int\limits_0^\infty\ee^{-\beta(H(y)-H(y-z))}\,dz\,=\,
y^{1-k}\int\limits_0^\infty\ee^{-\beta(H(y)-H(y-uy^{1-k}))}
\,du\,.
\label{into}
\qqq
We take $\,H(y)=ay^k+h(y)\,$ and assume that $\,h(y)\,$ is smooth
and that
\qq
h(y)\m|y|^{-k}\ \mathop{\rightarrow}\limits_{|y|\to\infty}\ 0\,,
\qquad\
(\partial_yh)(y)\m|y|^{-k+1}\
\mathop{\rightarrow}\limits_{|y|\to\infty}\ 0\,.
\qqq
First note that for $\,z>0$,
\qq
a\big(y^k-(y-z)^k\big)\,=\,a\sum\limits_{l=1}^k(-1)^{l+1}C_k^l\,
y^{k-l}z^l\,\geq\,3\epsilon z y^{k-1}+\epsilon z^k
\qqq
if $\,\epsilon>0\,$ is small enough. Next, for $\,0<z<\frac{1}{2}|y|\,$
and $\,|y|\,$ big enough,
\qq
|h(y)-h(y-z)|\,\leq\,z|\partial_yh(y-\vartheta z)|\,\leq\,\epsilon zy^{k-1}
\qqq
and for $\,z\geq\frac{1}{2}|y|$,
\qq
|h(y)-h(y-z)|\,\leq\,C\,+\,\epsilon z^k
\qqq
so that, altogether,
\qq
H(y)-H(y-z)\,\geq\,-C\,+\,\epsilon z y^{k-1}\,=\,-C+\epsilon u\,.
\qqq
for $\,z=uy^{1-k}$. \,It is also easy to see that for each $\,u>0$,
\qq
\lim\limits_{|y|\to\infty}\,\,[H(y)-H(y-uy^{1-k})]\,=\,aku\,.
\qqq
The estimate (\ref{estim}) follows then from the dominant convergence
theorem applied to the integral on the right hand side of Eq.\,(\ref{into}).

\appsection{}

We shall illustrate here the MFDT (\ref{mfd}) and its version (\ref{Fmfd}) 
in the frequency space on the simple example of the one-dimensional Langevin 
equation (\ref{1dLan}) on a circle with the Hamiltonian $\,H=0\,$ and 
a constant force $\,G$. \,In this case, Eq.\,(\ref{1dLan}) has, of course,
the explicit solution
\qq
x_t\,=\,x_0\,+\,Gt\,+\,\sqrt{%
2\beta^{-1}}
\,W(t)
\label{xt}
\qqq
for the standard Brownian motion $\,W(t)$. \,Since $\,x\,$ is the
angular variable, the above process possesses an invariant probability measure
with the constant density $\,\varrho_{\hspace{-0.01cm}_0}=\frac{1}{2\pi}$.
The corresponding current $\,j_{\hspace{-0.01cm}_0}=\frac{1}{2\pi}G\,$ 
is constant and so is the local mean velocity $\,\nu_{\hspace{-0.01cm}_0}=G$. 
\,The transition probabilities of the process have the form of the
shifted and periodized heat kernel 
\qq
&\displaystyle{P_t(x,dy)\ 
=\ \sqrt\frac{_\beta}{^{4\pi t}}\sum\limits_{n=\infty}^\infty
\ee^{-\frac{\beta|x-y+Gt+2\pi n|^2}{4t}}\,dy 
\ = \frac{_1}{^{2\pi}}\sum\limits_{n=-\infty}^\infty
\ee^{\m i n(x-y+Gt)-\beta^{-1}n^2t}\m dy}\cr
&\displaystyle{=\ \frac{_1}{^{2\pi}}\,\vartheta_3\big(\frac{_1}{^2}(x-y+Gt),\,
\ee^{-t/\beta}\big)\,,}
\qqq
where $\,\vartheta_3(z,q)\,$ is the Jacobi theta function \cite{GrRh}. 
\,For the (real) observables 
$\,O^a(x)=\sum\limits_n \hat{O}_n\ee^{i n x}\m$, \,one obtains easily
the equalities:
\qq
&&\CR^{ab}(t-s)\,=\,
\theta(t-s)\sum\limits_nO^a_n\m O^b_{-n}\,n^2\,
\ee^{\m(inG-\beta^{-1}n^2)(t-s)}\,,\cr
&&\CC^{ab}(t-s)\,=\,
\sum\limits_n O^a_n\m O^b_{-n}\,\ee^{\m in G(t-s)-
\beta^{-1}n^2|t-s|}\,,\cr
&&\CB^{ab}(t-s)\,=\,-i\m G\m\,\theta(t-s)\,
\sum\limits_n O^a_n\m O^b_{-n}\,n\,\,\ee^{\m(in G-\beta^{-1}n^2)(t-s)}\,.
\qqq
The result for the Fourier transforms (\ref{C}) and (\ref{RB}) 
follows immediately:
\qq
&&\hspace{-0.6cm}\hat\CR^{ab}(\omega)\,=\,\sum\limits_nO^a_n\m O^b_{-n}\,
\frac{n^2}{\beta^{-1}n^2-i\omega-inG}\,,\cr
&&\hspace{-0.6cm}\hat\CC^{ab}(\omega)\,=\,\sum\limits_nO^a_n\m O^b_{-n}\,
\frac{2\beta^{-1}n^2}{\beta^{-2}n^4+(\omega+nG)^2}\,,\cr
&&\hspace{-0.6cm}\hat\CB^{ab}(\omega)\,=\,-\sum\limits_nO^a_n\m O^b_{-n}\,
\frac{inG}{\beta^{-1}n^2-i\omega-inG}\,.
\qqq
The modified FDT (\ref{mfd}) or (\ref{Fmfd}) are clearly satisfied. Taking
$\,a=b$, \,one may define the factors
\qq
X^a(t-s)\,=\,\frac{
\CR^{aa}(t-s)}
{\beta\,\partial_s\,\CC^{aa}(t-s)}
\quad\ {\rm for}\quad\ t>s\,,
\qquad\qquad\hat X^a(\omega)\,=\,\frac{2\,{\rm Im}\,\CR^{aa}(\omega)}
{\beta\m\omega\m\,\hat\CC^{aa}(\omega)}
\qqq
that control the violation of the standard FDT \cite{CK,CKP}.
In particular, for $\,O^a(x)\,$ equal to $\,\sin(nx)\,$ or
$\,\cos(nx)\,$ with $\,n\not=0$, \,one obtains (replacing the superscript
$\,a\,$ by $\,n\,$ in this case):
\qq
X^n(t-s)\,=\,\Big[1+\frac{\beta G}{n}\,\tan\big(nG(t-s)\big)
\Big]^{\hspace{-0.04cm}-1}\,,
\qquad\hat X^n(\omega)\,=\,
\frac{\beta^{-2}n^4+\omega^2-n^2G^2}{\beta^{-2}n^4
+\omega^2+n^2G^2}\,.
\qqq
Note that $\,X^n(0)=1=\hat X^n(\infty)\,$ but that 
these factors are not necessarily positive. This is also true
for the ``effective temperatures''
\qq
T^a_{\rm eff}(t-s)\,=\,\beta^{-1}X^a(t-s)^{-1}\m,\qquad
\hat T^a_{\rm eff}(\omega)\,=\,\beta^{-1}\hat X^a(\omega)^{-1}\m.
\qqq
In particular, $\,T^n_{\rm eff}(\omega)\,$ is positive only in the region 
where $\,\omega^2>(n^2G^2-\beta^{-2}n^4)\,$ and in this region
it decreases with $\,\omega^2\,$ approaching for $\,\omega^2\to\infty\,$
the value $\,\beta^{-1}$. 
\vskip 0.1cm

The above calculations show that the equilibrium relation 
(\ref{EFDT}) between the response and correlation functions 
is strongly violated in the Langevin equation on a circle with 
a constant drift unless the drift vanishes. On the other hand,
the drift may be removed altogether by passing into the frame moving with 
constant velocity $\,\nu_{\hspace{-0.01cm}_0}=G$, \,see Eq.\,(\ref{xt}). 
\,This is captured by our MFDT (\ref{extend}). Indeed, the solutions 
of the advection equation (\ref{adv}) are
\qq
O^a(t,x)\,=\,\sum\limits_{n=-\infty}^{\infty}\hat{O}_n\,\ee^{in(x-Gt)}
\qqq
so that the Lagrangian-frame response and correlation functions are
\qq
\CR^{ab}_{_L}(t,s)\,=\,\CR^{ab}(t-s)|_{_{G=0}}\,,
\qquad\quad\CC^{ab}_{_L}(t,s)\,=\,\CC^{ab}(t-s)|_{_{G=0}}
\qqq
and the MFDT (\ref{extend}) takes the form of the equilibrium
FDT holding for $\,G=0$.

\end{document}